\documentclass[12pt]{spieman}  
\usepackage{amsmath,amsfonts,amssymb}
\usepackage{color,soul}
\usepackage{graphicx}
\usepackage{setspace}
\usepackage{tocloft}
\usepackage{siunitx}
\usepackage{makecell}
\usepackage{tabularx}
\usepackage{textcomp, gensymb}
\newcolumntype{L}[1]{>{\raggedright\let\newline\\\arraybackslash\hspace{0pt}}m{#1}} 
\usepackage[dvipsnames]{xcolor}

\title{Performance predictions and contrast limits for an ultraviolet high contrast imaging testbed}

\author[a]{Kyle Van Gorkom}
\author[a]{Ramya M.\ Anche}
\author[b]{Christopher B.\ Mendillo}
\author[c]{Jessica Gersh-Range}
\author[a]{Justin Hom}
\author[d]{Tyler D.\ Robinson}
\author[e]{Mamadou N'Diaye}
\author[f]{Nikole K.\ Lewis}
\author[g]{Bruce Macintosh}
\author[a]{Ewan S.\ Douglas}
\affil[a]{Steward Observatory, University of Arizona, 933 N Cherry Avenue, Tucson, AZ 85721, USA}
\affil[b]{Lowell Center for Space Science and Technology, University of Massachusetts Lowell, 600 Suffolk St. Suite 315, Lowell, MA 01854, USA}
\affil[c]{DM Telescopes LLC, Raleigh, NC, USA}
\affil[d]{Lunar \& Planetary Laboratory, University of Arizona, 1629 E University Blvd, Tucson, AZ 85721, USA }
\affil[e]{Université C{\^o}te d’Azur, Observatoire de la C{\^o}te d’Azur, CNRS, Laboratoire Lagrange, Nice, France}
\affil[f]{Department of Astronomy, Cornell University, Space Sciences Bldg, 404, 122 Sciences Dr, Ithaca, NY 14850, USA}
\affil[g]{UC Observatories, Astronomy \& Astrophysics Department, University of California Santa Cruz, CA 95064, USA}

\cftpagenumbersoff{figure}
\cftpagenumbersoff{table} 
\begin{document} 
\maketitle

\begin{abstract}

NASA's Habitable Worlds Observatory (HWO) concept and the 2020 Decadal Survey's recommendation to develop a large space telescope to ``detect and characterize Earth-like extrasolar planets" requires new starlight suppression technologies to probe a variety of biomarkers across multiple wavelengths. Broadband absorption due to ozone dominates Earth's spectrum in the mid-ultraviolet (200-300 nm) and can be detected with low spectral resolution. Despite the high value of direct ultraviolet (UV) exoplanet observations, high-contrast coronagraph demonstrations have yet to be performed in the UV. Typical coronagraph leakage sources such as wavefront error, surface scatter, polarization aberrations, and coronagraph mask quality all become more significant in the UV and threaten the viability of HWO to produce meaningful science in this regime. As a first step toward a demonstration of UV coronagraphy in a laboratory environment, we develop an end-to-end model to produce performance predictions and a contrast budget for a vacuum testbed operating at wavelengths from \SI{200}-\SI{400}{\nano \meter}. At \SI{300}{\nano \meter}, our model predicts testbed performance of ${\sim}3\times10^{-9}$ contrast in a narrow 2\% bandwidth and $\lessapprox10^{-8}$ in a 5\% bandwidth, dominated primarily by the chromatic residuals from surface errors on optics that are not conjugate to the pupil.
\end{abstract}


{\noindent \footnotesize\textbf{*}Kyle Van Gorkom,  \linkable{kvangorkom@arizona.edu} }


\section{Introduction}
\label{sect:intro}  
The next era of exoplanet astronomy will focus on searching for biosignatures, such as oxygen and methane, and will involve direct imaging and spectroscopy across ultraviolet to infrared wavelengths with an intended planet-to-star contrast of 1 in 10 billion ($10^{-10}$) \cite{}. For planets like Earth, abundant oxygen and ozone maintained in the presence of methane is thought to be a key sign of life \cite{meadows2018exoplanet,des2002remote,domagal2014abiotic}. The broadband absorption due to ozone is a significant signal that begins at around 350 nm and peaks below 300 nm, detectable even at low spectral resolution. Critically, the ozone Hartley (200--310\,nm) and Huggins (310--350\,nm) bands can indicate the presence of molecular oxygen even in weakly oxygenated atmospheres due to the non-linear relationship between ozone and oxygen concentrations \cite{2018ApJ...858L..14O}. Thus, the UV spectral range provides crucial access to ozone spectral features for oxygenated atmospheres \cite{2021AJ....161..150C,damiano2023reflected}. Additionally, the longer blue and violet wavelengths can reveal Rayleigh scattering effects, which helps constrain the atmospheric mean molecular mass, helping to better constrain the abundance of absorbers \cite{benneke2012atmospheric,feng2018characterizing,damiano2022reflected}.

UV coronagraphy also has applications in observing and characterizing actively accreting protoplanets and their circumplanetary disks (CPDs) \cite{zhu2015}. Few observations of these systems in the UV have been performed, with the only clear detection being PDS 70 b at 336 nm using HST-WFC3, without a coronagraph \cite{zhou2021}. Non-detections are likely contrast-limited, with WFC3 achieving a ${\sim}10^{-4}$ contrast at $<1''$ without a coronagraph. The \textit{Astro2020} Decadal Survey states that obtaining high spatial and low spectral resolution observations of CPDs will provide crucial insights into protoplanet masses, accretion rates, and planetary formation mechanisms.

Despite the potential value of direct UV observations of exoplanets, high-contrast coronagraph demonstrations in the UV have yet to be conducted. Currently, high-contrast imaging instruments in space, such as the Space Telescope Imaging Spectrograph (STIS) on the Hubble Space Telescope (HST), conduct coronagraphy within a broadband range of \SI{200}{\nano \meter} to \SI{1.2}{\micro \meter}, achieving a post-processed contrast of $10^{-6}-10^{-7}$\cite{debes2019, PM2024}. In the near future, the Nancy Grace Roman Space Telescope's coronagraph instrument is expected to reach a contrast of $10^{-9}$ within a 10\% bandpass filter centered at \SI{575}{\nano \meter} \cite{poberezhskiy2021roman,poberezhskiy2022roman,kasdin2020nancy}. As an alternative to an internal coronagraph, starshades have been proposed as a potential avenue through which HWO could realize high contrast imaging capability in the UV\cite{shaklan_2023,shaklan_2024}. Existing high contrast imaging testbeds (e.g., HCIT, DST, HiCAT, and THD2) \cite{noyes_2023,hicat_5,n2013high,serabyn_high-contrast_2013, thd2_wfc_2020, thd2_pol_2024} target $10^{-10}$ contrast in the \SI{500}-\SI{700}{\nano \meter} wavelength range and have yet to explore contrast performance at wavelengths $<$\SI{500}{\nano \meter}, leaving a significant technology gap.

Here, we present contrast budgeting and end-to-end performance modeling for a vacuum testbed designed for high contrast imaging in the UV, accounting for the dominant effects expected to set the contrast floor and incorporating as-built specifications for optical components where available or best estimates otherwise. In Section \ref{sec:hci-challenges}, we provide an overview of the physical terms expected to limit contrast performance in the UV and provide scaling laws where possible. Section \ref{sec:testbed-design} gives a description of the proposed configurations of the testbed for two coronagraph architectures, and Sections \ref{sec:prop_wfc}-\ref{sec:jitter-beamwalk} walk through the simulations and analysis of the individual terms considered. Section \ref{sec:contrast-budget} integrates all the terms for a combined contrast budget and end-to-end performance prediction.

\section{Challenges for high contrast imaging at ultraviolet wavelengths}
\label{sec:hci-challenges}

Recent white papers\cite{tuttle2024_hwo, roser_inprep} and the final report of the Coronagraph Technology Roadmap Working Group\cite{ctr_2024} have highlighted a number of challenges expected for HWO to achieve high contrast at UV wavelengths, including tighter requirements on wavefront sensing and control, higher sensitivity to contamination and scattered light, stricter requirements on coating performance and polarization aberrations, and lower photon flux due to a combination of both lower system throughput and dimmer source magnitude in the UV (which presents challenges from both an SNR and WFS\&C perspective). In this section, we present a subset of the terms expected to dominate the contrast budget at UV wavelengths and quantify the effect size when considering wavelengths 2-3x shorter than visible. The scope of the analysis here and in the numerical treatment that follows is focused on instrument-level performance and in general does not consider the performance limits set by telescope optics or observatory-level effects.

\subsection{Contrast sensitivity to surface and reflectivity errors}

Uncorrected surface and reflectivity aberrations inject speckles in the focal plane at separations dictated by their spatial frequency content. In the small phase regime, the electric field from a sinusoidal surface aberration  of amplitude $\alpha$ with a spatial frequency of $n$ cycles per aperture $D$ can be approximated (ignoring the aperture function) as
\begin{equation}
    E \approx 1 + i \alpha \frac{4\pi}{\lambda} \cos \left( 2 \pi \frac{n}{D} x \right)
\end{equation} In a bandwidth $B=\Delta \lambda/ \lambda_0$, the post-coronagraph contrast at the corresponding $\pm n \lambda/D$ focal plane separation will be approximately given by (Equation \ref{eqn:contrast_bandwidth_nocorr} in Appendix \ref{sec:appA})
\begin{equation}
    C \approx \alpha^2 \frac{4 \pi^2}{\lambda_0^2} \left[ \frac{1}{1-\left( \frac{B}{2} \right)^2} \right]
\end{equation}

For an uncorrected aberration of a given amplitude, the contrast scales as $\lambda_0^{-2}$. In a general optical system, the vast majority of optical surfaces will be located in planes that are not conjugate to the pupil; in this case, wavefront aberrations cycle between phase and amplitude at propagation distances characterized by the Talbot length\cite{Talbot_1836}
\begin{equation}\label{eqn:z_talbot}
    z_T = \frac{2} {\lambda (n/D)^2}.
\end{equation}
At every $z_T/4$ distance, phase is completely converted into amplitude and vice versa. The wavelength-dependence of this effect means that surface and amplitude aberrations on out-of-pupil optics limit the quality of the wavefront correction as bandwidth increases\cite{Shaklan06, pueyo_polychromatic_2007, mazoyer_pueyo}. Note that Talbot lengths scale inversely with wavelength, so the Talbot length for an aberration of a fixed spatial frequency is longer in the UV compared to visible wavelengths. To first order, however, the contrast floor after DM correction of phase and amplitude aberrations with Talbot-induced chromaticity is independent of the central wavelength and depends only on the bandwidth\cite{mazoyer_pueyo} (Equation \ref{eqn:contrast_bandwidth_approx}). Higher order terms---including cross-terms from aberrations and the deformable mirror (DM) correction at multiple spatial frequencies (e.g., frequency-folding\cite{giveon_2006})---do, however, show a strong dependence on the central wavelength. With a single DM nulling speckles in a half-sided dark hole (DH), the contrast limit at $n \lambda/D$  arising from the second-order term of a sinusoidal phase error of amplitude $\beta$ at $n/2D$ cycles/aperture has the form (Equation \ref{eqn:contrast_secondorder})
\begin{equation}
    C_{\beta} = \beta^4 4 \pi^4 \left[ \frac{1}{\lambda_0^4} f_1(B) + \frac{\pi^2 z^2 \left(\frac{n}{D} \right)^4}{\lambda_0^2} f_2(B) \right],
\end{equation}
where $f_1(B)$ and $f_2(B)$ are functions purely of the bandwidth, and $z$ is the distance from the aberrated optic to the pupil plane. This contrast term is dominated by the $\lambda_0^{-4}$ dependence and implies a $16-81\times$ larger contribution to the overall contrast at UV wavelengths compared to visible. Pueyo and Kasdin 2007\cite{pueyo_polychromatic_2007} showed that this analysis can be extended to calculate the broadband contrast floor to an arbitrary order in the Taylor expansion of the complex exponential. Because typical optical surface errors represent a larger fraction of a wavelength in the UV compared to visible, second- and higher-order terms are expected to play a larger role in setting the achievable UV contrast floor, although the relative amplitude of these terms depends on surface quality and optical design choices. Closed-form expressions for the bandwidth-integrated contrast due to the Talbot effect for a handful of simple cases are derived in Appendix \ref{sec:appA}.

\subsection{Deformable mirror electronics}\label{sec:dm_electronics}

Quantization of the commands sent to the deformable mirror (DM) actuators to the least significant bit (LSB) sets a floor on the precision of the electric field correction and, therefore, the achievable contrast. Ruane et al.\ 2020\cite{ruane_quant_2020} derive an expression for the quantization contrast floor for a Boston Micromachines Corp. (BMC) Kilo-C MEMS device, assuming Gaussian influence functions
\begin{equation}\label{eqn:quant}
    C_\mathrm{quantization}(\alpha) = \frac{16 \pi}{3 n_\mathrm{act}^2} \left( \frac{h_\mathrm{min}} {\lambda} \right)^2 \pi^2 \omega^4 e^{-(\alpha/\alpha_\mathrm{infl})^2},
\end{equation}
where $\alpha$ is the angular separation in the focal plane, $n_\mathrm{act}$ is the number of actuators across the pupil (after the Lyot stop), $h_\mathrm{min}$ is the minimum actuator step size, $\omega=d/p$ is the ratio of the influence function radius $d$ to the actuator pitch $p$, and $\alpha_\mathrm{infl} = \frac{\sqrt{2} \lambda} {2 \pi d}$. For small separations ($\alpha \ll \alpha_\mathrm{infl}$) where quantization has the largest influence, the contrast $C_\mathrm{quantization} \propto \lambda^{-2}$. At a fixed, non-zero angular separation, the contrast scales more modestly, with a form given by
\begin{equation}
    C_\mathrm{quantization}(n, \lambda) \propto \left( \frac{\lambda_0} {\lambda} \right)^2 \exp \left\{- \left[ \frac{n}{n_\mathrm{infl}} \right]^2 \left[ \left(\frac{\lambda_0}{\lambda} \right)^2 - 1 \right] \right\},
\end{equation}
where $n$ is the angular separation in $\lambda/D$, $n_\mathrm{infl}=\frac{\sqrt{2} n_\mathrm{act}}{2 \pi \omega \Gamma}$, and $\Gamma$ is the Lyot stop fraction. At a separation of $6 \lambda/D$ at $\lambda=\SI{200}{\nano \meter}$ and assuming typical values for a coronagraph system with a Kilo-C DM\cite{ruane_quant_2020}, the quantization-limited contrast is a factor of ${\sim}6$ larger compared to $\lambda_0=\SI{600}{\nano \meter}$ at the same separation ($2 \lambda_0/D$). DM electronics for a UV coronagraph will require a $4-9\times$ lower quantization limit compared to visible to achieve an equivalent contrast.

This analysis can also be extended to estimate the effect of actuator noise (e.g., due to voltage noise from the electronics at temporal frequencies higher than the high-order (HO) sensing bandwidth) on the contrast. Equation \ref{eqn:quant} assumes quantization errors that are uniformly distributed over a $V_\mathrm{LSB}$ range (the voltage step corresponding to the LSB), with voltage variance $\sigma_{\Delta v, \mathrm{LSB}}^2 = V^2_\mathrm{LSB} / 12$ and $h_\mathrm{min} = g V_\mathrm{LSB}$ (where $g$ is the actuator gain in surface displacement per volt). If we instead assume an independent, identically-distributed Gaussian noise term on each actuator, the variance is simply $\sigma_{\Delta v, \mathrm{Gaussian}}^2 = V^2_\mathrm{noise}$ with a corresponding actuator displacement $h_\mathrm{noise} = g V_\mathrm{noise}$. Replacing $h_\mathrm{min}$ in Equation \ref{eqn:quant} with $h_\mathrm{noise}$, the contrast floor due to this term is given by
\begin{equation}\label{eqn:dm_noise}
    C_\mathrm{noise}(\alpha) = 64 \frac{\pi}{n_\mathrm{act}^2} \left( \frac{h_\mathrm{noise}} {\lambda} \right)^2 \pi^2 \omega^4 e^{-(\alpha/\alpha_\mathrm{infl})^2},
\end{equation}
which implies that electronics noise must be a factor of $\sqrt{12}$ smaller than the LSB to avoid limiting the contrast. This expression has the same wavelength and angular separation dependence as the quantization term. 

Note that DM quantization places a limit on the correction of the coherent electric field, while DM electronics noise injects an additional incoherent intensity term. DM dynamics at slower rates (e.g., drift or creep) within the sensing bandwidth can be corrected with either a dark zone maintenance (DZM) algorithm\cite{miller_ldfc,manojkumar_2024,redmond_2024} or low-order wavefront sensing and control (LOWFS)\cite{Guyon_2009,singh_llowfs_2014,Singh_2015, mendillo_2023}.

\subsection{Scattering and contamination}

There are multiple causes of scattered light in a coronagraphic system. Defects in the focal plane mask\cite{sidick.2014}, scratches and dust particles on the optical surfaces\cite{dohlen.2008,balasubramanian.2009}, and surface roughness of the reflective optics\cite{church.1990,stover.1995,elson.1995,dittman.2006,harvey.2007,nelson.2007,harvey.2009,krywonos.2011,harvey.2012} all contribute to the scattered light background. Surface roughness can be quantified statistically; it is distributed across the surface and isotropic on large scales. The bidirectional scattering distribution function (BSDF) for each surface is defined based on the surface power spectral density (PSD).

\begin{equation}
  \label{eqn:bsdf}
  B\!S\!D\!F = \frac{16\pi^2}{\lambda^4}cos(\theta_i)cos(\theta_s) P\!S\!D 
\end{equation}

Here $\lambda$ is the wavelength of light, $\theta_i$ is the incident angle of a ray and $\theta_s$ is the scattering angle of the ray. The BSDF has a strong ($1/\lambda^4$) dependence on wavelength, making it much more significant at shorter wavelengths. The scattered light background at \SI{300}{\nano \meter} is 16 times greater than for visible corongraphs operating at \SI{600}{\nano \meter}.

Contaminants on optical surfaces are another source of scattering. Performance modeling for HST treated dust on the primary mirror as a population of small absorbers, the primary effect of which (apart from a throughput hit) was to diffractively scatter light out to the wings of the point spread function (PSF)\cite{Hasan_1995}. End-to-end diffraction simulations of the Gemini Planet Imager (GPI) modeled dust on each surface in a similar fashion and developed predictions for contrast degradation as a function of cleanliness level\cite{marois_2008}. More modeling is required to understand the implications of contamination effects for high contrast imaging  at $10^{-10}$ contrast in the UV\cite{ctr_2024, tuttle2024_hwo}.

\subsection{Jitter and beam walk}

Dynamic line of sight errors---induced by either telescope pointing errors or sources of vibration in an instrument or testbed environment---create leakage through the coronagraph that can limit the contrast\cite{ruane_2017}. The magnitude of the effect depends strongly on the coronagraph architecture. For a vector vortex coronagraph (VVC)\cite{mawet_annular_2005,foo_optical_2005,mawet_optical_2009,ruane_vortex_2018,Serabyn19}, the requirement on pointing stability to achieve a given contrast scales as $\lambda$\cite{ruane_2017}, while coronagraphs that primarily rely on pupil-plane masks are largely insensitive to jitter.

A second-order term that arises from jitter is a translation of the beam on intermediate optics that generates dynamic high spatial frequency content in the wavefront error seen by the coronagraph, an effect known as beam walk\cite{Guyon_2005, mendillo_tolerances}. Even if the jitter is sensed and corrected by a LOWFS loop, beam walk will be present on optical surfaces upstream of the fast steering mirror (FSM). The net result is an incoherent floor of speckles within the dark zone. For a jitter RMS $\sigma$ that creates beam walk on an optic a distance $z$ from the pupil with a surface error of amplitude $\alpha$ at $n/D$ cycles/aperture, the incoherent contrast floor is given by
\begin{equation}\label{eqn:beamwalk}
    C_\mathrm{beam walk} = \alpha^2 \frac{2 \pi^2}{\lambda^2} \left[1 - e^{-2\pi^2 \left( z \frac{n}{D} \sigma \right)^2} \right],
\end{equation}
which has a $\lambda^{-2}$ dependence on wavelength. For a given jitter environment and optical system, the contrast floor due to beam walk is expected to be $4-9\times$ larger in the UV compared to the visible. See Appendix \ref{sec:beamwalk_deriv} for a derivation of this expression.

\subsection{Polarization aberrations}\label{sec:polarization-aberrations-challenges}
\begin{figure}[!ht]
    \centering
    \includegraphics[width=1\linewidth]{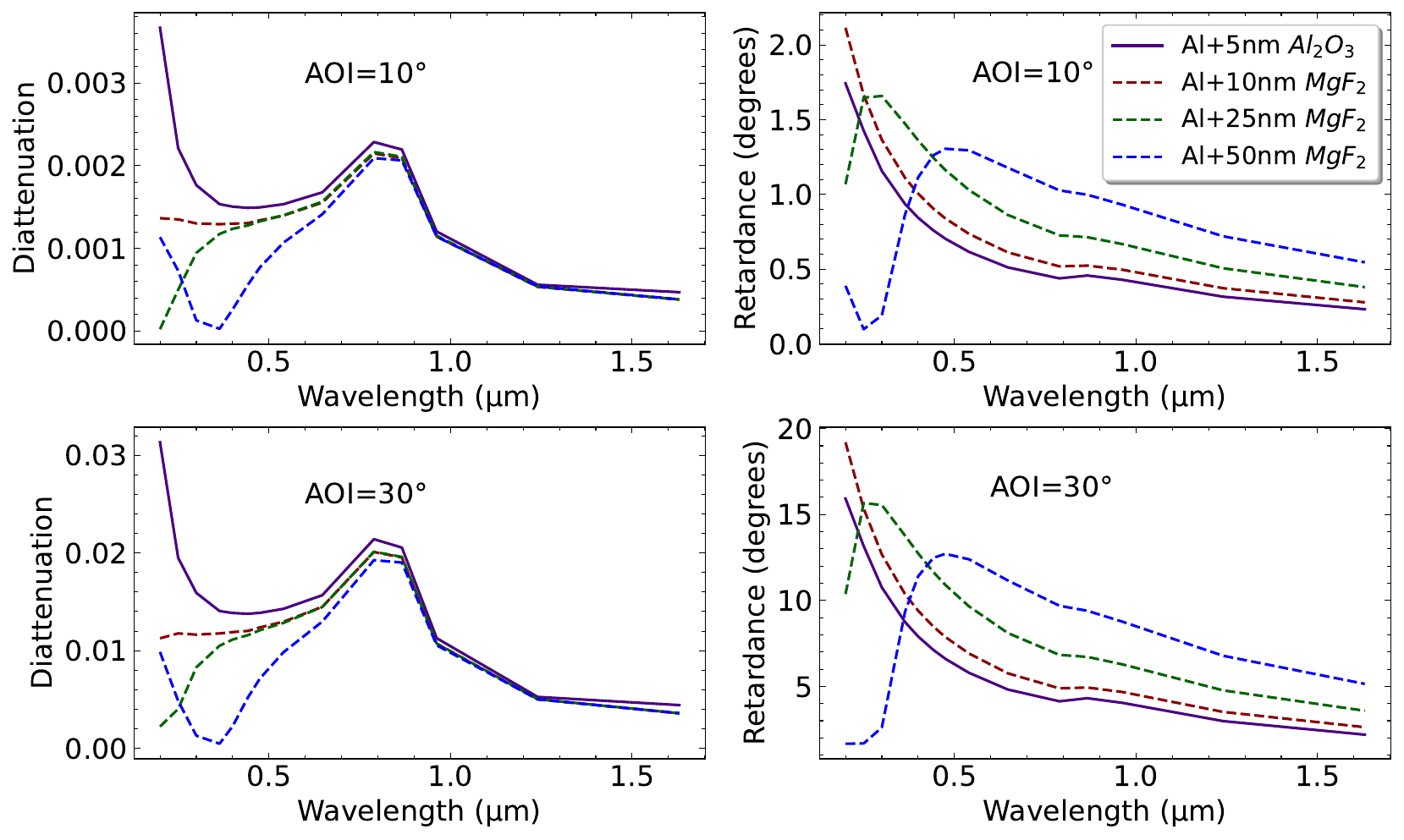}
    \caption{Diattenuation and retardance estimated for different thicknesses of the protective layer $\rm MgF_2$ on aluminum.}
    \label{fig:diat-retardance}
\end{figure}

Polarization aberrations (described by diattenuation and retardance) are a function of the refractive indices of the optical coatings and angles of incidence on the surfaces in an instrument's optical train\cite{Breckinridge_2015,anche_2023}. For a single-layer oxidized aluminum coating, diattenuation and retardance decrease with the wavelength, following the refractive index variation as shown in Figure \ref{fig:diat-retardance}. The aberrations are most sensitive to the angle of incidence (AOI); increasing the AOI by a factor of 3 increases the diattenuation and retardance by a factor of 10. As $\rm MgF_2$ is the most commonly-used protective layer for the aluminum coating, we compare the diattenuation and retardance for different thicknesses of $\rm MgF_2$ to understand the scaling with wavelength. Adding the $\rm MgF_2$ layer reduces the diattenuation aberrations considerably at the cost of an increase in the retardance aberration. Comparing the retardance curves for different  $\rm MgF_2$ thicknesses in Figure \ref{fig:diat-retardance}, adding a 50nm layer of $\rm MgF_2$ almost doubles the retardance aberrations compared to oxidized aluminum. At the same time, the reduction in the diattenuation is not significant.

In addition to the changes in diattenuation and retardance caused by $\rm MgF_2$, another polarization-dependent effect that could influence the achievable contrast is the birefringence of $\rm MgF_2$, which is approximately 0.012-0.013 at UV wavelengths \cite{dodge1984refractive, tuttle2024_hwo}. We note, however, that although the MgF$_2$ crystal exhibits high birefringence at ultraviolet wavelengths, a thin film of MgF$_2$ deposited at room temperature, well below the crystal's melting point, is expected to form nanocrystals in random orientations, resulting in no effective birefringence \cite{rodriguez2017self}.

The non-uniformity of coating deposition is another significant effect to consider at UV wavelengths. For the Gemini telescope, a coating thickness non-uniformity of 15\% (or 1.3 nm) peak-to-valley variation was measured over a 4 cm × 4 cm substrate \cite{schneider2016gemini}. A recent study simulated coating non-uniformity for ground-based large-segmented mirror telescopes (GSMTs) and found that the contrast degradation for a second-order perfect coronagraph with 10-20\% variation in thickness of the protective coating is on the order of $2\times 10^{-8}$ at visible wavelengths. This degradation is below the adaptive optics residuals for GSMTs \cite{ashcraft2025}, but may prove to be a significant challenge at UV wavelengths for HWO.

\section{Testbed design and optical layouts}
\label{sec:testbed-design}
\begin{figure}
    \centering
    \includegraphics[width=1\linewidth]{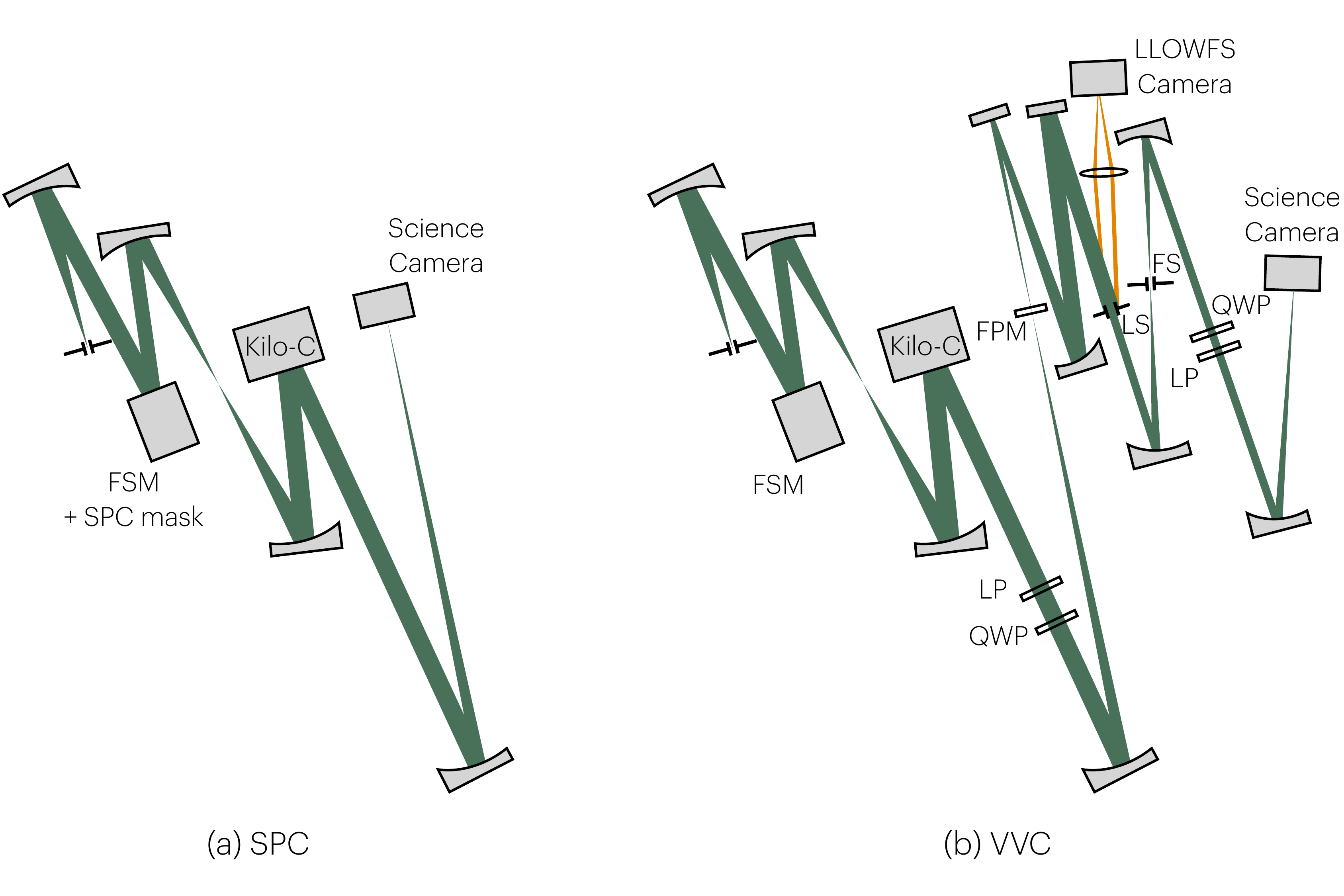}
    \caption{Optical layouts of SCoOB for (a) the SPC configuration and (b) the VVC configuration.}
    \label{fig:layouts}
\end{figure}

The Space Coronagraph Optical Bench (SCoOB)\cite{jaren,mespie} was optimized for visible-wavelength high contrast imaging but designed to accommodate UV wavelengths (e.g., $\rm Al+MgF_2$ coatings for high throughput down to \SI{200}{\nano \meter}) with relatively minor modifications. In its current configuration, SCoOB comprises a BMC Kilo-C 34x34 deformable mirror, a VVC from Beam Co.\ with polarization filtering, a reflective Lyot stop to enable Lyot LOWFS\cite{singh_llowfs_2014,Singh_2015,mendillo_2023}, and a Sony IMX571 sensor in a vacuum-compatible form-factor from Neutralino Space Ventures. The testbed is installed in a thermal vacuum chamber (TVAC) from Rydberg Vacuum Systems capable of $<10^{-6}$ Torr and mounted on a Minus K passive vibration isolation platform for mechanical isolation. The testbed has achieved $2\times10^{-9}$ contrast in a narrow bandwidth at visible wavelengths\cite{vangorkom_2024}, and a number of modifications are in progress that are expected to improve on this performance.

We consider two configurations of the testbed for the UV, a layout consisting of a shaped pupil coronagraph (SPC)\cite{kasdin2003,vanderbei2003a,vanderbei2003b} and one with a UV-optimized charge-6 VVC (see Figure \ref{fig:layouts}). The SPC is one of simplest coronagraphs, and it requires a small number of optics, thereby minimizing the amount of scattering from optical surface roughness and the throughput penalty due to low reflectance in the UV. It is also inherently achromatic and relatively robust to polarization, jitter, wavefront aberrations, misalignments, and the presence of dust. The SPC configuration comprises only an SPC mask in the entrance pupil (co-located with the FSM), a relay to the DM, and a science camera in the f/48 focal plane immediately following the DM. The SPC mask is assumed to be a silicon wafer with either black silicon (BSi) or carbon nanotubes (CNT) to define the mask and a bare aluminum coating on the reflective portion. The SPC mask is designed to create a $<10^{-10}$ contrast DH from $6-14\lambda/D$. This DH was selected as a starting point based on the results of a design survey that examined how the inner working angle (IWA) and outer working angle (OWA) affect the clear area of the SPC mask (which is a proxy for the core throughput)\cite{Gersh-Range_2025}. The maximum IWA in this survey, 6$\lambda/D$, was based off of a target star list for HWO that identifies 164 stars with a minimum IWA of 6.3$\lambda/D$ for a telescope diameter of \SI{6}{\meter} and a wavelength of \SI{300}{\nano \meter}. The OWA has less of an impact on the clear area and accessible design space for SPCs. The survey considered OWA up to 15$\lambda/D$, but individual point designs have been produced for OWA as high as 60$\lambda/D$. We do not include a field stop (FS) in the model here, but one could be easily accommodated by placing it in the f/48 beam and shifting the science camera to the next focal plane (i.e., the nominal FS position in the VVC configuration). A reflective FS would also allow LOWFS for wavefront stabilization.

\begin{figure}
    \centering
    \includegraphics[width=\linewidth]{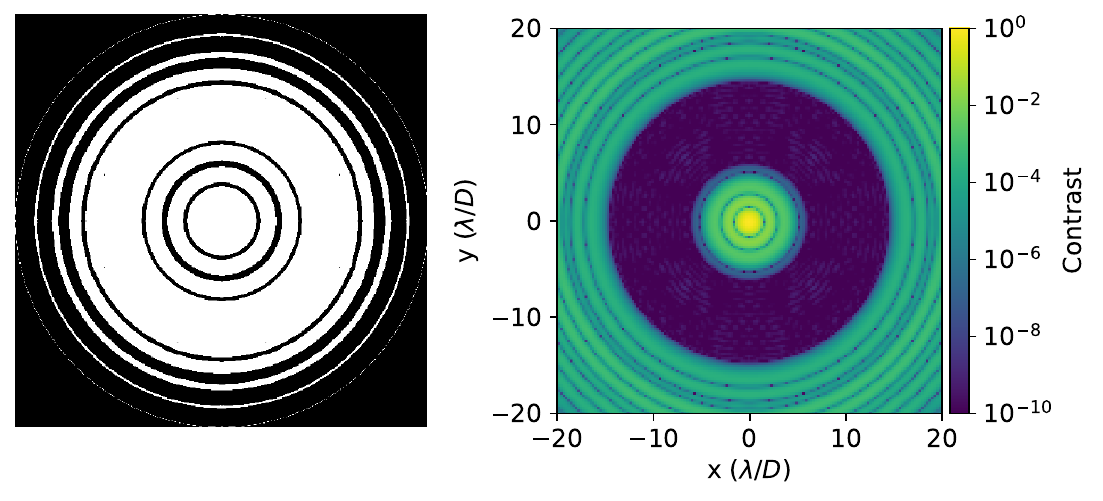}
    \caption{Left: Design of the SPC mask assumed here. Right: The $6-14\lambda/D$ dark hole created by the SPC in the absence of any sources of error.}
    \label{fig:spc_nominal}
\end{figure}

The VVC comprises a spatially-varying halfwave plate in a focal plane with a Lyot stop in a downstream pupil plane. The optical axis of the halfwave plate varies azimuthally to accumulate an integer-multiple number of $2 \pi$ radians of phase in one complete rotation, which remaps on-axis light to outside the geometrical pupil after propagation to a downstream pupil plane. This integer multiple is referred to as the charge of the device. VVCs have been shown to be capable of producing high contrast in the laboratory environment, including the Decadal Survey Testbed (DST)\cite{riggs_high_2021}, the High Contrast Spectroscopy Testbed for Segmented Telescopes (HCST)\cite{llop-sayson_high-contrast_2020}, and the High Contrast Imaging Testbed (HCIT)\cite{serabyn_high-contrast_2013, prada_high-contrast_2019,ruane_experimental_2020}. A charge-6 VVC has a ${\sim}3\lambda/D$ IWA, defined by the 50\% throughput separation\cite{ruane_2017}. We select a $3-10\lambda/D$ DH in the VVC configuration to explore a region of separation space that is complementary to the $6-14\lambda/D$ SPC DH.

The proposed VVC configuration is identical to the current visible layout of SCoOB, with the VVC and polarization-filtering optics swapped out for their UV-optimized equivalents. A linear polarizer (LP) and quarter wave plate (QWP) are placed in the collimated beam downstream of the DM to create circular polarization, and a second QWP and LP are placed in the collimated space after the field stop to filter the leakage term from retardance errors on the VVC. The entrance pupil is a reflective stop, again fabricated with either BSi or CNT for low reflectivity. The Lyot stop is nominally 95\% the pupil diameter. As with the SPC, the field stop is not included in these simulations but can be included should it prove necessary in the laboratory environment. Both SPC and VVC configurations feature a single DM, so only half-sided dark holes (DH) are considered here.

The source is fiber-fed into the testbed and re-imaged onto a \SI{2}{\micro \meter}-diameter micro-fabricated pinhole\cite{jenkins_2023}. The beam is collimated by a \SI{146.2}{\milli \meter}-focal length off-axis parabola (OAP) and then stopped to \SI{6.8}{\milli \meter} at the entrance pupil. As a fraction of a resolution element, the pinhole diameter is $0.47 - 0.23\lambda/D$ from $\si{200}-\SI{400}{\nano \meter}$. The DM is modeled as a 100\% yield, \SI{1.5}{\micro \meter}-stroke, bare aluminum-coated BMC Kilo-C device with 952 actuators on a \SI{300}{\micro \meter} pitch. The pupil re-imaged onto the DM is ${\sim}\SI{9.27}{\milli \meter}$, for a beam footprint approximately 31 actuators across. We assume 16-bit control electronics, which sets a floor on the minimum actuator step size (see Section \ref{sec:prop_wfc}). The science camera in both configurations is assumed to be the same Sony IMX571 sensor used in the visible (\SI{3.76}{\micro \meter} pixel pitch), with a fluorescent coating to increase the quantum efficiency (QE) in the UV\cite{acton_report}.

\section{Diffraction model and wavefront control}
\label{sec:prop_wfc}

To predict the performance of the testbed and identify the dominant terms considered in Section \ref{sec:hci-challenges}, we built a numerical model of SCoOB in \textsc{HCIPy}\cite{hcipy} and performed end-to-end Fresnel diffraction simulations incorporating surface and reflectivity errors (with ${\sim}\SI{13.3}{\micro \meter / pixel}$ sampling), DM actuator quantization\cite{ruane_quant_2020} and electronics noise, jitter and beamwalk, and polarization aberrations. \textsc{HCIPy} natively supports the simulation of polarized electric fields, so we also directly incorporate polarization optics (including retardance errors and finite extinction) in the Fresnel simulations. An incoherent floor due to scattering from surface roughness not captured in the diffraction model is separately computed following the approach adopted for PICTURE-C performance modeling\cite{mendillo_tolerances} and added to focal plane image produced by the Fresnel diffraction model.

We model the output of the \SI{2}{\micro \meter}-diameter source pinhole as a Gaussian field and propagate it analytically to the first OAP, which collimates and relays the beam to the entrance pupil stop. The primary effect of this Gaussian beam is to add a low-order amplitude aberration at the input of the system.

\subsection{Wavefront sensing and control}

We simulate two types of focal-plane wavefront sensing and control (FPWFS\&C): an initial pupil-plane wavefront flattening routine, and dark hole digging via electric field conjugation (EFC) \cite{efc}. In practice, the first step is often performed by sensing the pupil-plane phase via phase retrieval\cite{thurman,vangorkom_2021} at the science camera and closed-loop correction with the DM. We note that the ideal pupil plane for wavefront flattening depends on the coronagraph mask. In the SPC configuration, the system exit pupil suffices; in the VVC configuration, the aberrations should be corrected in the virtual pupil upstream of the FPM in order to avoid compensating for aberrations accumulated downstream of the FPM that do not contribute to leakage through the mask. Since our simulations give us direct access to the pupil-plane phase, we simulate this by projecting this phase map onto the DM influence functions to compute the actuator commands that will flatten the wavefront. In the case of the SPC, we correct the exit pupil phase in the absence of the SPC mask in order to avoid phase wrapping problems across the mask boundaries. For the VVC configuration, this same plane is the virtual pupil upstream of the FPM. The DM command that flattens the wavefront creates the initial state of the system for EFC.

To estimate the contrast floor after EFC, we run monochromatic simulations that target a half-sided DH. The DM Jacobian is computed by running the forward propagation model with pairs of small (${\sim}$\SI{1}{\nano \meter}) pokes on each actuator and measuring the corresponding change in the focal-plane electric field. When digging, we re-linearize by recomputing the Jacobian every 2-5 iterations and employ a form of beta-bumping\cite{seo_beta_bumping}, which has been found to improve performance when EFC iterations start to stagnate. Rather than simulate pairwise probing for electric field estimation, we assume perfect knowledge of the focal-plane electric field to distinguish sensing errors from control limits. The outcomes of these EFC simulations are then evaluated in 2\%, 5\%, and 10\% bandwidths.

\begin{figure}
    \centering
    \includegraphics[width=1\linewidth]{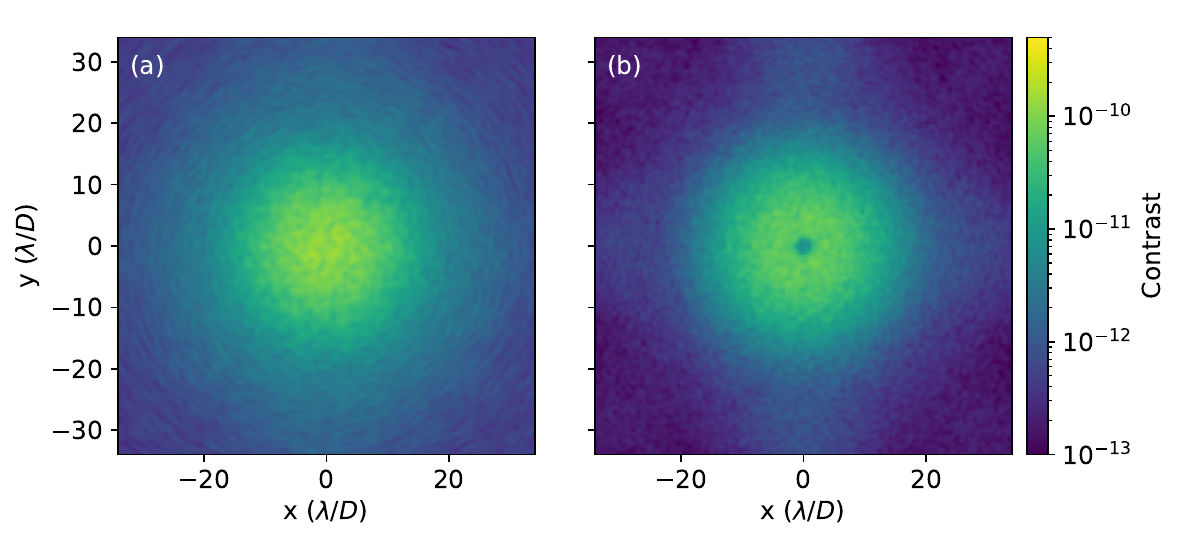}
    \caption{Simulated contrast residuals due to \SI{1.2}{\milli \volt} (\SI{3.5}{\pico \meter}) Gaussian noise on each actuator at \SI{300}{\nano \meter} for (a) the SPC configuration and (b) the VVC configuration. The contrast floor set by this term is isolated by subtracting the difference of the post-EFC exit pupil fields with and without actuator noise and incoherently averaging the residual focal plane intensities for 50 noise realizations. The low throughput of the charge-6 VVC at small separations\cite{ruane_2017} creates the observed dip in contrast at the center of the field, while the SPC throughput is constant as a function of separation.}
    \label{fig:actnoise_sim}
\end{figure}

The resolution limit due to the 16-bit DM electronics is directly included in the closed-loop EFC simulations. We assume a Kilo-C DM with \SI{1.5}{\micro \meter} stroke over a \SI{195}{\volt} range and 16-bit electronics. Given the quadratic stroke-voltage relationship for these devices, the minimum achievable step size can be traded for total stroke by running at a low voltage bias. At a 40\% voltage bias, the actuator gain is ${\sim}\SI{2.9}{\nano \meter / V}$ for a minimum step size of ${\sim}\SI{8.7}{\pico \meter}$. In our numerical simulations, we convert the requested actuator displacement commands to bits, truncate, and then convert back to displacement. We analytically estimate the contrast floor due to this quantization following Equation \ref{eqn:quant}. At \SI{200}{\nano \meter}---the shortest central wavelength considered here and the most sensitive to DM quantization---we estimate the corresponding contrast floor to be $8\times10^{-11}$. Our DM electronics have a typical measured voltage noise of ${\sim}\SI{1.2}{\milli V}$ (or 40\% the LSB)\cite{haughwout_2023}. Plugging this into Equation \ref{eqn:dm_noise} at \SI{200}{\nano \meter} yields a contrast floor of $2\times10^{-10}$. An example of the incoherent contrast contribution from actuator noise is shown in Figure \ref{fig:actnoise_sim}. Apart from jitter and DM electronics noise, these simulations do not include any form of system dynamics or control with a LOWFS loop.

\section{Surface and reflectivity errors}
\label{sec:surface-reflectivity-errors}
\begin{figure}
    \centering
    \includegraphics[width=1\linewidth]{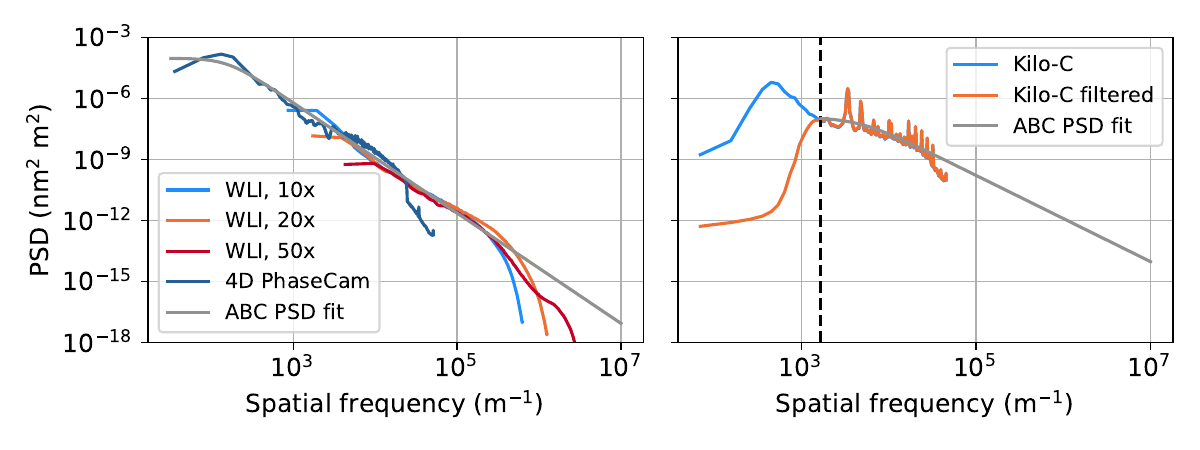}
    \caption{Left: PSD measurements of the SCoOB OAPs and a PSD fit. Here, $b=\SI{154}{\meter}^{-1}\approx2.5$ cycles/aperture, $c=2.7$, and the RMS surface is \SI{4.7}{\nano \meter}.  Right: PSD measurements of a Kilo-C device. The dashed line indicates the highest spatial frequency correctable by the DM, $1/(2d) \approx \SI{1667}{\meter}^{-1}$, where $d$ is the \SI{300}{\micro \meter} actuator pitch. Here, $b=\SI{4513}{\meter}^{-1}\approx41$ cycles/aperture, $c=2.1$, and the RMS surface is \SI{8.7}{\nano \meter}. The interferometric measurement of the DM is interpolated onto the Fresnel model sampling and used directly in the diffraction simulation. The ABC PSD fit is used to extrapolate to high spatial frequncies for the surface roughness scattering model. Note that each measured PSD curve shows a fall-off toward higher spatial frequencies, indicative of the instrument transfer functions. These higher spatial frequencies are excluded from the PSD fit to avoid a bias that would under-represent high spatial frequency content.}
    \label{fig:surface_psds}
\end{figure}

Surface and reflectivity errors at each optic are characterized by a two-dimensional power spectral density (PSD) of the form
\begin{equation}\label{eqn:psd}
    \mathrm{PSD}(k) = \frac{a}{1 + (k/b)^{c}},
\end{equation}
which captures the aberration amplitude at each spatial frequency $k$. This particular form of the PSD is commonly used to characterize optical surface quality for high contrast imaging applications\cite{mendillo_tolerances, krist_cgi}. The parameter $b$ sets the knee frequency, the frequency below which the PSD flattens out, and $c$ is the power-law slope beyond the knee frequency. The parameter $a$ sets the maximum power of the PSD, but in practice is often ignored in favor of an overall surface RMS normalization.

To estimate the PSD parameters of our as-built optics, we measured a subset of our off-axis parabolas (OAPs) with a 4D PhaseCam 6000 interferometer, with a resolution of \SI{9.8}{\micro \meter / pixel} and sub-nanometer surface repeatability. Following the approach in Krist et al.\ 2023\cite{krist_cgi}, we subtracted out Zernike polynomials up to Z11 (following the Noll ordering convention) and applied a radial Tukey window prior to the compution of the radial PSD. To sample the PSD out to higher spatial frequencies, we followed up with additional measurements with a white light coherence scanning interferometer (WLI), the Zygo NewView 8300, with 10x-50x microscope objectives that enabled resolutions down to \SI{0.16}{\micro \meter / pixel}. On each OAP, we measured 5 distinct portions of the surface. To remove alignment errors and the local curvature of the OAP, we subtracted off piston, tip, tilt, and focus from each WLI measurement prior to windowing and computing the PSDs.

The averaged radial PSDs computed from these measurements and a best-fit model are shown in Figure \ref{fig:surface_psds}. Synthetic surfaces are created from the model PSD by generating a two-dimensional representation of the radial PSD, scaling to amplitude, applying random phases, and taking the Fourier transform. The resulting surface map is scaled to the desired RMS over the clear aperture.

\begin{figure}
    \centering
    \includegraphics[width=1\linewidth]{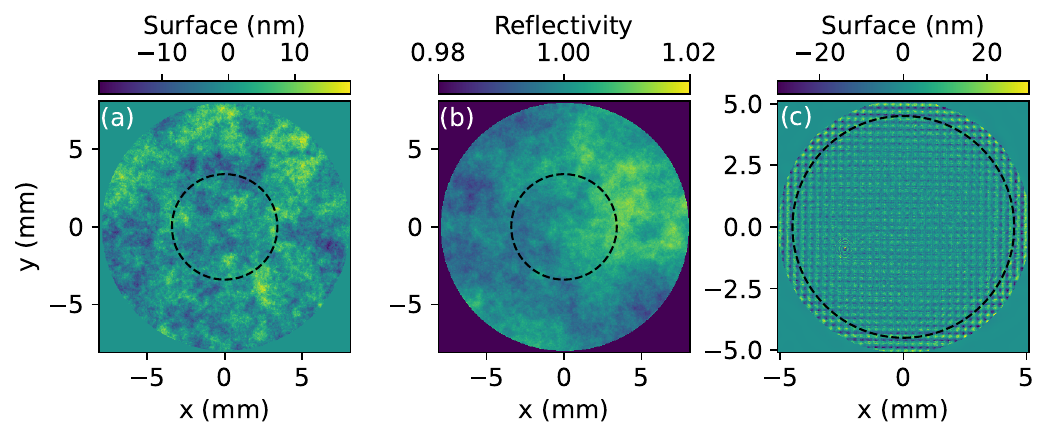}
    \caption{Left: Synthetic surface height errors generated via PSD for the OAP immediately after the entrance pupil. Middle: Synthetic reflectivity maps generated via PSD for the same OAP. Right: Surface map for a Kilo-C DM measured via interferometry. The dashed circles mark the nominal beam size on each optic.}
    \label{fig:surfaces}
\end{figure}

Reflectivity maps are generated in a similar fashion. In the absence of direct measurements of reflectivity variations, we assume a PSD with knee frequency $b=10$ cycles/meter, $c=2.65$, and RMS = 0.5\%\cite{mendillo_tolerances}. Examples of a synthetic surface and reflectivity map are shown in Figure \ref{fig:surfaces}.

The Fresnel model incorporates random draws of these surface and reflectivity error maps at each optic in the optical train. The exception to this approach is the DM surface. The DM surface is taken into account by directly including an interferometric measurement of a Kilo-C DM with a sampling of \SI{9.96}{\micro \meter / pixel} (or about 30 pixels/actuator) and interpolating onto the model sampling. See Figures \ref{fig:surface_psds} and \ref{fig:surfaces}.

\begin{figure}
    \centering
    \includegraphics[width=1\linewidth]{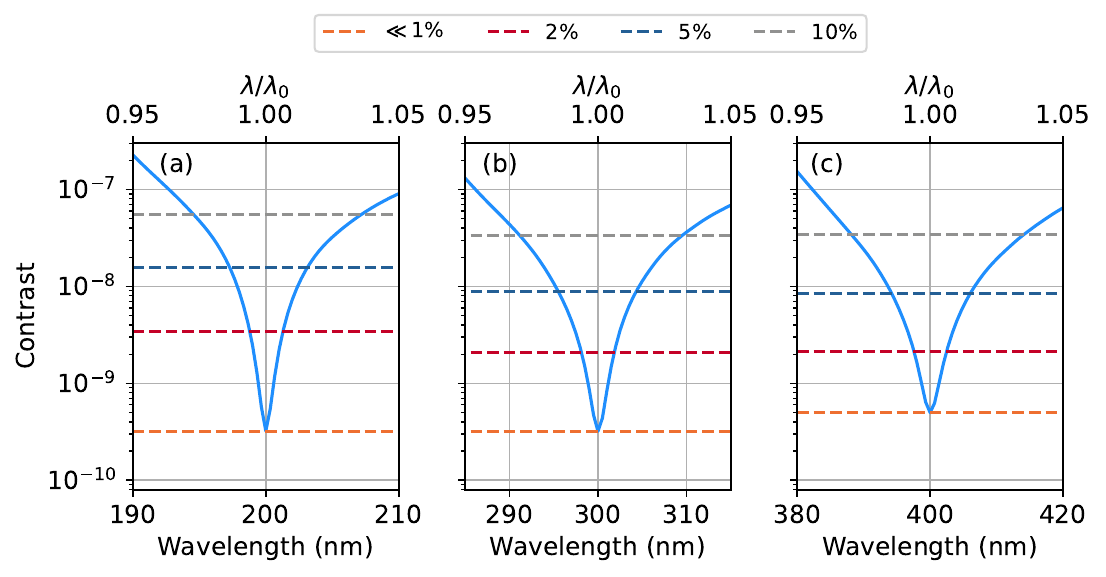}
    \caption{SPC contrast in a 6-14$\lambda_0/D$ DH for central wavelengths (a) $\lambda_0=\SI{200}{\nano \meter}$, (b) $\lambda_0=\SI{300}{\nano \meter}$, and (c) $\lambda_0=\SI{400}{\nano \meter}$. Solid lines show the spectral dependence of the contrast, and dashed lines indicate the mean contrast in the DH in bandwidths up to 10\%. Only surface and reflectivity errors are included in these simulations.}
    \label{fig:spc_efc}
\end{figure}

\begin{figure}
    \centering
    \includegraphics[width=1\linewidth]{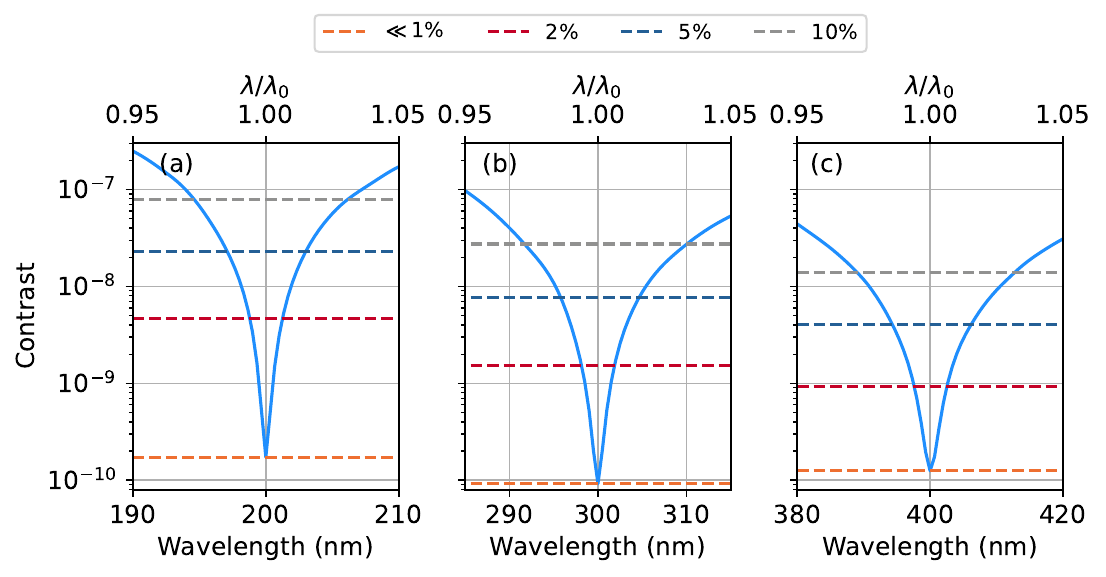}
    \caption{ VVC contrast in a 3-10$\lambda_0/D$ DH for central wavelengths (a) $\lambda_0=\SI{200}{\nano \meter}$, (b) $\lambda_0=\SI{300}{\nano \meter}$, and (c) $\lambda_0=\SI{400}{\nano \meter}$. Solid lines show the spectral dependence of the contrast, and dashed lines indicate the mean contrast in the DH in bandwidths up to 10\%. Only surface and reflectivity errors are included in these simulations.}
    \label{fig:vvc_efc}
\end{figure}

Figures \ref{fig:spc_efc} and \ref{fig:vvc_efc} show the simulated contrast floor set by these surface and reflectivity errors after performing EFC for the SPC and VVC configurations, at wavelengths from \SI{200} to \SI{400}{\nano \meter} and bandwidths (BWs) up to 10\%. 

A full treatment of the scattering from each surface requires sampling up to the $1/\lambda$ spatial frequency. Our Fresnel model has a sampling of \SI{13.3}{\micro \meter / pixel}, so it can capture spatial frequencies up to ${\sim}4\times10^{4}\ \si{\meter}^{-1}$. For $\lambda=\SI{200}{\nano \meter}$, however, $1/\lambda$ corresponds to a spatial frequency of $5\times10^{6}$ $\si{\meter}^{-1}$, which requires ${\sim} 10^5$ pixels across the aperture to capture. This sampling would require very large memory usage with an end-to-end diffraction model, so scattering at the highest spatial frequencies is simulated independently, as described in the next section.

\section{Scattering from surface roughness}
\label{sec:scattering}

The analysis of scattering from surface micro-roughness performed in this work will largely follow the method described in Dittman 2006\cite{dittman.2006}, which is based on the work done by Church \& Takacs 1990\cite{church.1990} and Stover 1995\cite{stover.1995}. This particular method was used in a previous study of optical specifications for the PICTURE-C balloon mission\cite{mendillo_tolerances}. The scattering angle $\theta_s$ relates to the spatial frequency $k$ of the surface error by the grating equation $\beta=sin(\theta_s)=k \lambda$, here using the convention that $\theta_s$ is measured relative to the nominal specular reflection angle of the ray. With a maximum scattering angle of $\theta_s = 90\degree$ ($\beta=1$), the corresponding maximum spatial frequency required to fully capture all possible scattering angles is $k_{max} = 1/\lambda$, or more simply, the minimum spatial period is equal to the wavelength of the light. 

The BSDF is defined in Equation \ref{eqn:bsdf}, with the PSD defined in Equation~\ref{eqn:psd} and the given $a$,$b$,$c$ values (Figure~\ref{fig:surface_psds}) are used for each optic. Using the definition of $\beta$ above, the probability $P$ (or fraction) of light scattering into an angular range $\beta_1$ to $\beta_2$ is given by Equation~\ref{eqn:sprob}.
\begin{equation}
  \label{eqn:sprob}
  P = \int_{\beta_1}^{\beta_2} B\!S\!D\!F \beta \,\mathrm{d}\beta
\end{equation}
And, the total integrated scatter (TIS) is given by Equation~\ref{eqn:tis}.
\begin{equation}
  \label{eqn:tis}
  T\!I\!S = \int_{0}^{1} B\!S\!D\!F \beta \,\mathrm{d}\beta
\end{equation}

\begin{figure}
  \begin{center}
    \includegraphics[width=0.98\textwidth]{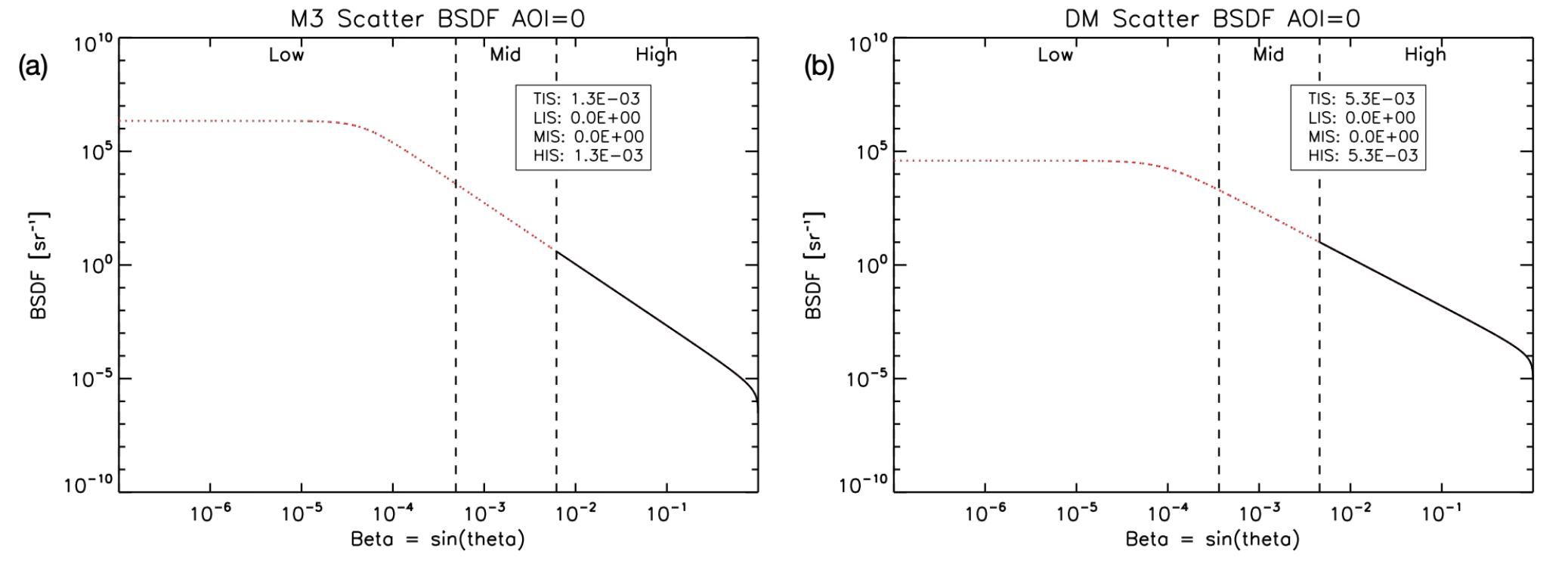}
  \end{center}
  \caption[bsdf]{\label{fig:bsdf}The bidirectional scattering distribution functions (BSDFs) of a system OAP (left) and the BMC Kilo DM (right) are calculated from the surface PSDs using Equation \ref{eqn:bsdf} at $\lambda = 300$~nm. The base BSDF is shown as the red dashed line and the Nyquist-truncated function used for scattering is shown in black. The higher RMS surface error and lower power-law index of the DM surface (Figure \ref{fig:surface_psds}) create a total integrated scatter (TIS) that is $\sim4$ times larger than the other optics. LIS, MIS, and HIS correspond to the integrated scatter at low-, mid-, and high-spatial frequencies.} 
\end{figure}

A random scattering angle can be drawn from the BSDF by first calculating the scattering cumulative distribution function (CDF).
\begin{equation}
  \label{eqn:cdf}
  C\!D\!F(\beta) = \int_{0}^{\beta} B\!S\!D\!F \beta' \,\mathrm{d}\beta'
\end{equation}
A random number generator is used to select a random probability from 0 to 1. The CDF is then inverted ($\beta$ as a function of probability) and solved (interpolated) for $\beta$ at the generated probability. If the probability is greater than the TIS, then the ray does not scatter. This will produce of distribution of $\beta$ values that follows the BSDF. To model only spatial frequencies not captured by the Fresnel model, the BSDF of each optic is set to zero below the Nyquist frequency of the Fresnel grid sampling as shown in Figure~\ref{fig:bsdf}.

\begin{figure}
  \begin{center}
    \includegraphics[width=0.98\textwidth]{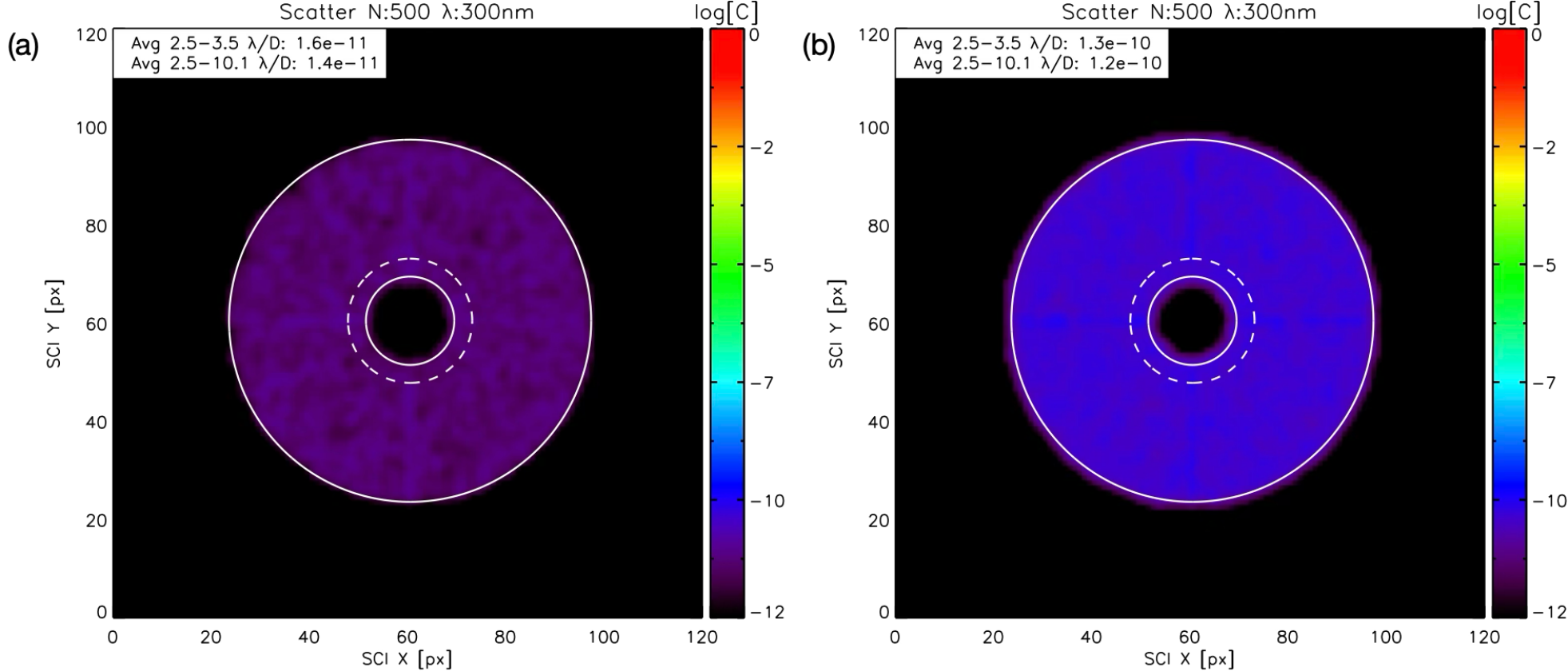}
  \end{center}
  \caption[scattering]{\label{fig:scatter}Simulations of leakage due to surface roughness scattering for the SPC (left) and VVC (right) configurations (see Figure~\ref{fig:layouts}). The more minimal SPC configuration has 6 scattering optics, while the VVC configuration has 17. The SPC scattering background is estimated to be $1.4\times10^{-11}$ and the VVC background is an order of magnitude larger at $1.2\times10^{-10}$}
\end{figure}

Once the filtered CDF of each optic has been constructed, a scattering raytrace can be used to simulate the multiple scattering of rays through the system. This analysis uses an efficient pixel-scattering model\cite{mendillo_tolerances}, which calculates the probability at each surface that a ray will scatter into a given pixel in the dark zone. This is the combined probability that the ray will scatter into the angular range subtended by the pixel and not scatter (change direction) at any of the downstream surfaces. The remaining probability, that the ray did not scatter directly into the pixel, is carried over to the next surface after first scattering in a random direction drawn from the CDF of the current surface. This process is repeated for each pixel in the dark zone to create the scattering contrast maps shown for the SPC and VVC configurations of SCoOB in Figure~\ref{fig:scatter}. The VVC setup is predicted to have a scattering background of $1.2\times10^{-10}$ compared to $1.4\times10^{-11}$ for the SPC. This order of magnitude increase is caused by the larger number of scattering optics in the VVC setup, 17 compared to only 6 for the SPC. 

\section{Polarization aberrations and retardance errors}
\label{sec:polarization-aberrations}

\begin{figure}[!ht]
    \centering
    \includegraphics[width=1\linewidth]{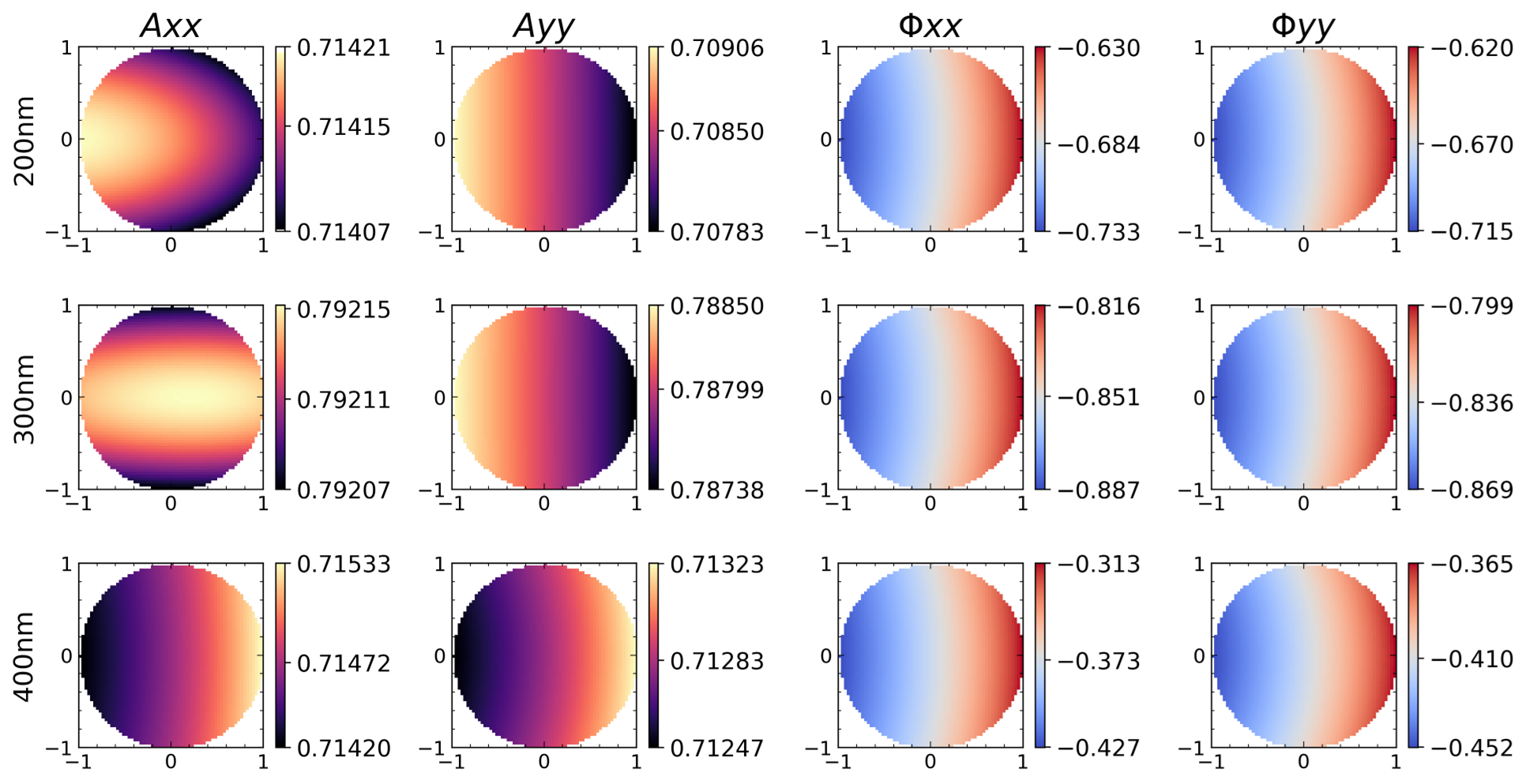}
    \caption{The Jones amplitude (left) and phase (right) maps of the diagonal terms estimated at the exit pupil of the SPC configuration.}
    \label{fig:jones-spc-diagonal}
\end{figure}

\begin{figure}[!ht]
    \centering
    \includegraphics[width=1\linewidth]{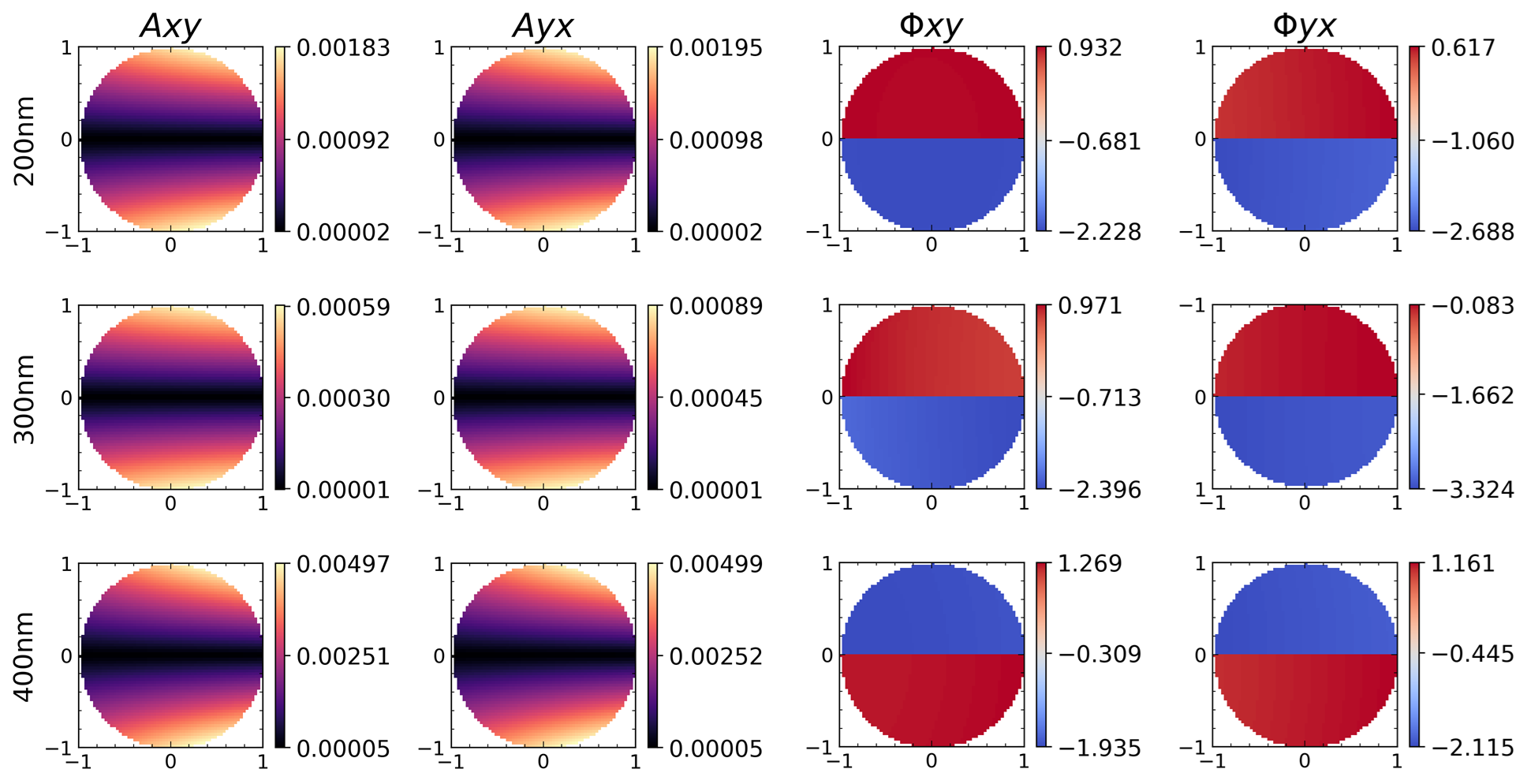}
    \caption{The Jones amplitude (left) and phase (right) maps of the cross terms estimated at the exit pupil of the SPC configuration.}
    \label{fig:jones-spc-crossed}
\end{figure}

To estimate the effect of polarization aberrations on system performance, Jones matrices are generally estimated at the exit pupil of the optical system \cite{Breckinridge_2015,anche2024,anche_2023}. We perform ray tracing using the Zemax OpticStudio API and estimate the Jones matrices using polarization ray tracing \cite{anche_2023}. We assume a protected aluminum ($\rm Al$+185nm $\rm MgF_2$) coating for all the OAPs and fold mirrors and oxidized aluminum ($\rm Al$+5nm $\rm Al_2$O$_3$) on the FSM and DM. We assume no effective birefringence for the thin film of $\rm MgF_2$ considered here (see Section \ref{sec:polarization-aberrations-challenges} for discussion). The Jones matrix at the SPC exit pupil (EP) (or the virtual pupil upstream of the FPM in the case of the VVC configuration) is shown in Figures \ref{fig:jones-spc-diagonal} and \ref{fig:jones-spc-crossed}. The Jones matrix at the VVC configuration EP is shown in Figures \ref{fig:jones-diag-vvc} and \ref{fig:jones-crossed-vvc}.

The diagonal terms of the Jones matrix (\textit{Axx}, \textit{Ayy}, \textit{$\phi xx$}, and \textit{$\phi yy$}) represent the amplitude and phase for the orthogonal $X$ ($E_x=1$,$E_y=0$) and $Y$ ($E_x=0$,$E_y=1$) polarized light as it propagates through the system. Figure \ref {fig:jones-spc-diagonal} shows these diagonal terms for three central wavelengths: \si{200}, \si{300}, and \SI{400}{\nano \meter}. The SPC EP amplitude terms vary $\sim$0.05-0.1\% over the pupil, dominated by diattenuation defocus, tilt, and piston. Similarly, the phase terms are dominated by retardance piston and tilt and vary by 0.07-0.1 radians ($\sim$2-6nm).
The off-diagonal terms (\textit{Axy}, \textit{Ayx}, \textit{$\phi xy$}, and \textit{$\phi yx$}) represent the amplitude and phase of \textit{X}-polarized light converted to \textit{Y}-polarized light and vice versa. The peak values of the cross components \textit{Axy} and \textit{Ayx} are 0.2\%, 0.05\%, and 0.5\%, respectively for \SI{200}, \SI{300}, and \SI{400}{\nano \meter} as shown in Figure \ref{fig:jones-spc-crossed}. The off-diagonal phase components \textit{$\phi xy$} and \textit{$\phi yx$} exhibit a $\pi$ phase shift when the corresponding cross-term amplitude passes through zero.


\subsection{Coupling between polarization aberrations and surface errors}\label{sec:pol_shear}

\begin{figure}[!ht]
    \centering
    \includegraphics[width=1\linewidth]{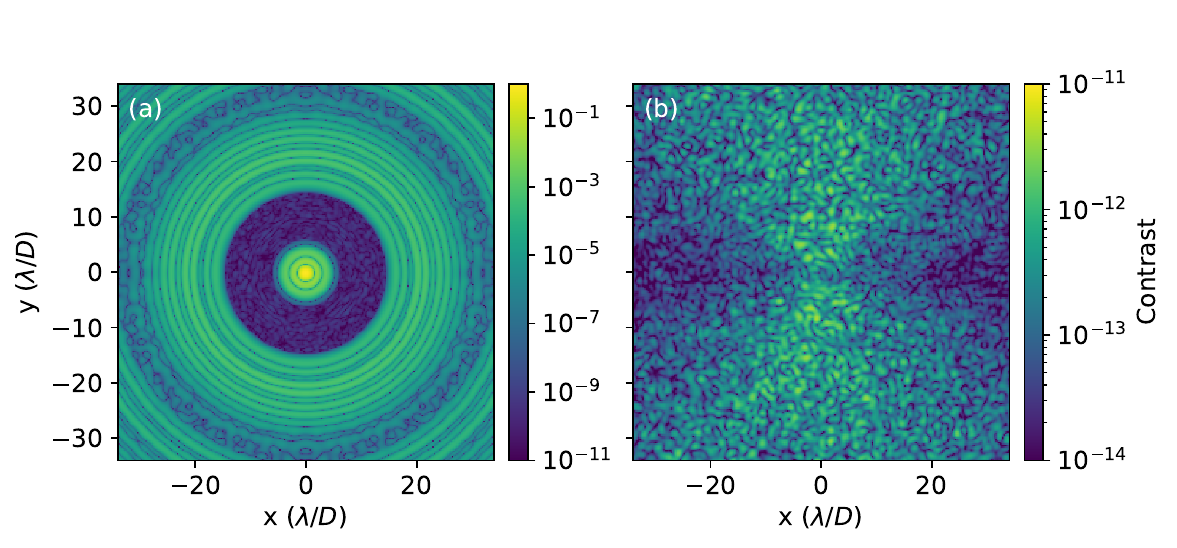}
    \caption{Simulations of the speckles induced by polarization aberrations for (a) the SPC and (b) the VVC configurations at \SI{300}{\nano \meter}, assuming perfect filtering (in the case of the VVC) and a perfect correction of the phase and amplitude aberrations common to the orthogonal polarization states at the DM. Because of the relatively small amplitude of the polarization aberrations in our testbed and the robustness of these coronagraph masks against low-order aberrations, the contrast floor is dominated by the polarization shear effect described in the text.}
    \label{fig:polabs_dhs}
\end{figure}

We integrate the polarization aberrations with our Fresnel diffraction model by multiplying the propagated vector electric field with a spatial Jones matrix at each optical surface. By including these Jones matrices at intermediate planes (rather than just in a pupil plane), the diffraction model captures a differential shear of the beam on each surface in the orthogonal polarization states. As a consequence of this shear, the orthogonal fields sample a slightly shifted portion of each optic, resulting in high spatial-frequency aberrations that are not common between the polarization states and cannot be corrected (akin to a static beamwalk term). As polarization aberrations are not simply tip/tilt, the coupling between surface errors and polarization aberrations is not strictly a shear, but the net result is the same: an incoherent speckle field out to larger separations than predicted by typically low-order polarization aberrations. Note that because this polarization-surface interaction generates non-common aberrations at high spatial frequencies, even coronagraphs that are robust against low-order aberrations (e.g., a high-charge VVC) are sensitive to this effect. A simulation of this term is shown in Figure \ref{fig:polabs_dhs}.

\subsection{Retardance errors and leakage}

Imperfect retardance on the vector vortex waveplate allows a fraction of the on-axis starlight to leak through the coronagraph mask. This leakage can be mitigated by polarization filtering with circular polarizers upstream and downstream of the focal plane mask. The leakage term with imperfect filtering is given by\cite{ruane_2019}
\begin{equation}
    L \approx \frac{1}{\gamma^2} \left( \frac{\epsilon_V^2} {4} + \epsilon_Q^2 \cos\left(2 \chi \right) \right),
\end{equation}
where $\gamma^2$ is linear polarizer extinction ratio, $\epsilon_V^2$ is the retardance error on the vortex mask, $\epsilon_Q^2$ is the retardance error on the quarter-wave plates, and $\chi$ is the vortex fast axis orientation. Based on measurements of our component-level performance at visible wavelengths, we adopt $\gamma^2=10^4$ and $\epsilon_V=\epsilon_Q=2^{\circ}$. At the worst value of $\chi$, this yields a leakage of $1.5\times10^{-7}$ on-axis, which is $<10^{-10}$ averaged over a $3-10\lambda/D$ DH.

\section{Mask fabrication errors}
\label{sec:mask-fabrication}

The coronagraph mask design process is typically idealized in the sense that it does not account for manufacturing errors, which can be difficult to predict in advance. To ensure that the manufactured masks achieve the designed contrast despite deviations from their optimized designs, we use an error budget to allocate the maximum acceptable contribution from each error source, with the specific allocations determined by a combination of simulation and prior experience. The important considerations include the reflectance (of both the bare aluminum and the black silicon for a reflective SPC), the surface figure error, the fidelity with which the shaped pupil pattern is manufactured, and the effects of mask defects. Using the requirements and lessons learned for Roman's BSi SPC masks as a guide, we have developed conservative error estimates for a UV coronagraph.

\subsection{Reflectance}

For a reflective SPC, both the specular and diffuse reflectance are restricted to prevent contrast degradation, but they are analyzed in different ways because they affect the contrast very differently. For a BSi mask, the specular reflections off of the aluminum and black silicon create PSFs that are incoherent, while the diffuse reflection off the black silicon leads to an incoherent near-uniform illumination of the dark hole\cite{balaMilestone1}. We estimate the impact of these reflections in the UV by following analysis methods used for Roman and assuming reflectances based on measurements of BSi and CNT masks.

The specular component of the reflection from the BSi masks fabricated for the Roman coronagraph was measured to be $<10^{-7}$, and the hemispherical reflectance $<0.3\%$ over visible wavelengths (\SI{400}{\nano \meter}-\SI{1}{\micro \meter})\cite{Balasubramanian_2019}. Measurements of the hemispherical reflectance of both BSi and CNTs tend to show an increase in reflectance toward UV wavelengths\cite{butler_2010,georgiev_2019}, although the spectral performance depends on the details of the fabrication process and can likely be optimized for better UV performance. To account for the uncertainties in the UV reflectance values, we adopt numbers that are $2\times$ worse than the upper limits of the Roman coronagraph masks, specular reflectance $R_s=2\times10^{-7}$ and hemispherical reflectance $R_d=0.6\%$. 

\begin{figure}
    \centering
    \includegraphics[width=1\linewidth]{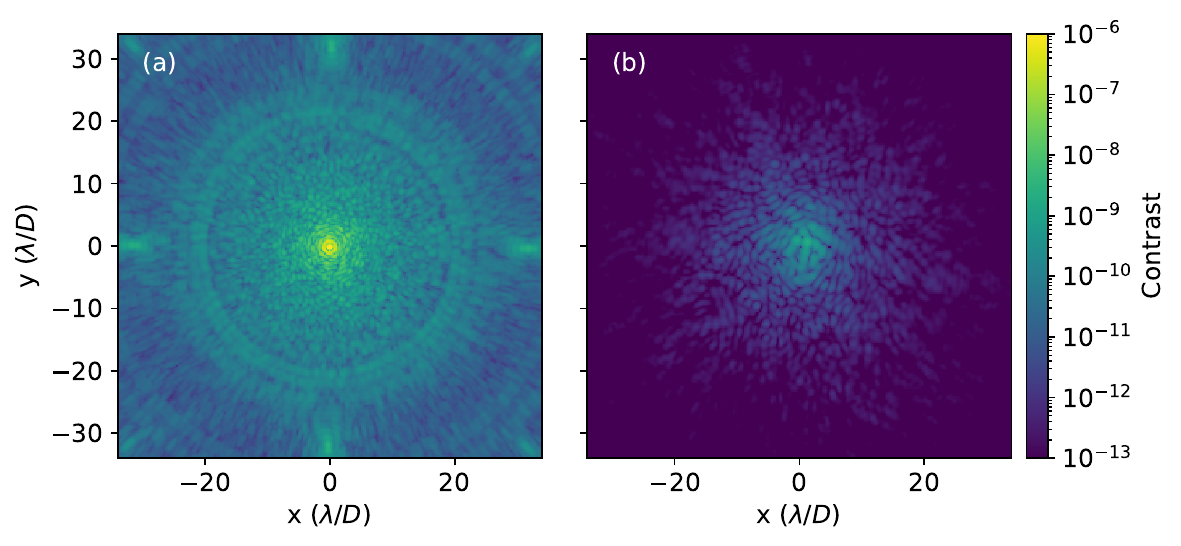}
    \caption{Specular contribution in the focal plane from the BSi/CNT portion of the simulated masks for (a) the SPC mask and (b) the entrance pupil stop for the VVC in a 5\% bandwidth centered at \SI{300}{\nano \meter}, assuming a $2\times10^{-7}$ specular reflectance.}
    \label{fig:mask_specular}
\end{figure}

Following the procedure in Balasubramian et al. 2015\cite{Balasubramanian_2015}, we calculate the contrast floor due to the specular component by assigning the ``non-reflective'' portions of the masks (shaped pupil in the case of the SPC configuration and circular stop for the VVC configuration) the specular reflectivity $R_s$ in the Fresnel model and propagate this term independently through the optical system. The focal-plane contrast from this term is added incoherently to the post-EFC image. The contrast maps for the specular mask reflection are shown for the two configurations in Figure \ref{fig:mask_specular}. At \SI{300}{\nano \meter}, this term contributes an incoherent intensity at ${\sim}10^{-10}$ contrast for the SPC and a ${\sim}10^{-11}$ contrast for the VVC.

We estimate the diffuse contribution from the mask by assuming that the BSi/CNT surface acts as a Lambertian scatterer and scatters a $(\lambda/D)^2 R_d$ fraction of the incident beam into small angles corresponding to the dark hole in the focal plane\cite{Balasubramanian_2015}. To compute the corresponding contribution to the contrast, we weight this term by the ratio of the total power illuminating the masked region of the optic to the power in the reflective region and normalize to the summed intensity within a $(\lambda/D)^2$ area for the nominal off-axis (non-coronagraphic) PSF. This approximation assumes the scattering is uniform over the DH and ignores interactions of this scattered light with optics downstream of the mask. At \SI{300}{\nano \meter}, we estimate this term to be at the $2\times10^{-10}$ contrast level for the SPC mask and $8\times10^{-11}$ for the pupil stop in the VVC configuration.

\subsection{Defects}
\begin{figure}
    \centering
    \includegraphics[width=1\linewidth]{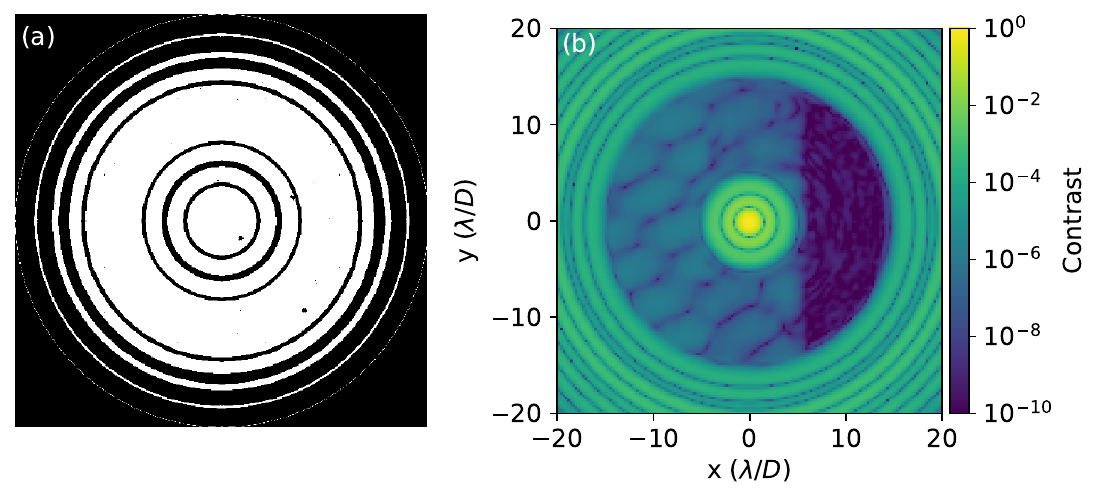}
    \caption{Left: Realization of an SPC mask with simulated amplitude defects. Right: Monochromatic dark hole after EFC to compensate for SPC defects.}
    \label{fig:spc_defects}
\end{figure}

The primary defects identified on the Roman shaped pupil masks were found to be isolated, low-reflectivity regions on the otherwise reflective portions of the masks\cite{riggs2024romanmasks,Balasubramanian_2015,Balasubramanian_2019}. The majority of these defects were relatively small (${\sim}\SI{100}{\micro \meter}^2$) with a small handful exceeding $\SI{1000}{\micro \meter}^2$. To estimate the effect of similar defects on our shaped pupil mask design, we generated two populations of defects and applied them to the ideal masks as amplitude errors. For 10 realizations at each central wavelength considered here, we randomly distributed 20 single-pixel ($\SI{169}{\micro \meter}^2$) defects and 3 larger defects each comprising a ${\sim}$20-pixel area ($\SI{3380}{\micro \meter}^2$). After EFC, the majority of cases converged to ${<}5\times10^{-10}$ monochromatic contrast. See Figure \ref{fig:spc_defects} for a realization of an SPC mask with defects and the corresponding post-EFC DH. The EFC correction for these defects is chromatic, and we report the quality of the correction in bandwidths up to 10\% in Table \ref{tab:contrast_spc}.

\subsection{Surface errors}
Surface errors and reflectivity variations (apart from the defects described above) on the SPC masks were generated from the PSD of Equation \ref{eqn:psd}, assuming the same PSD parameters as those used for the OAP surfaces. This is likely an underestimate of the low-order aberrations expected on the mask, but---considering that the SPC mask is in a conjugate pupil and that low-order errors are within the bandwidth and stroke of the DM---the omission of these terms is not expected to affect estimates of system performance.

\section{Jitter and beam walk}\label{sec:jitter-beamwalk}

Jitter is modeled by injecting tip/tilt errors at the front end of the system and running multiple realization of the diffraction model with tip/tilt values pulled from a Gaussian distribution with a ${\sim}6\times10^{-3} \lambda/D$ standard deviation at $\lambda=\SI{300}{\nano \meter}$. This value is based on jitter measured at the SCoOB science camera under vacuum conditions\cite{vangorkom_2024}, in the absence of LOWFS or other forms of jitter control. By injecting jitter as tip/tilt in the system, beam walk on intermediate (out-of-pupil) optics that injects high-frequency aberrations into the system can be captured in the model.

\begin{figure}
    \centering
    \includegraphics[width=1\linewidth]{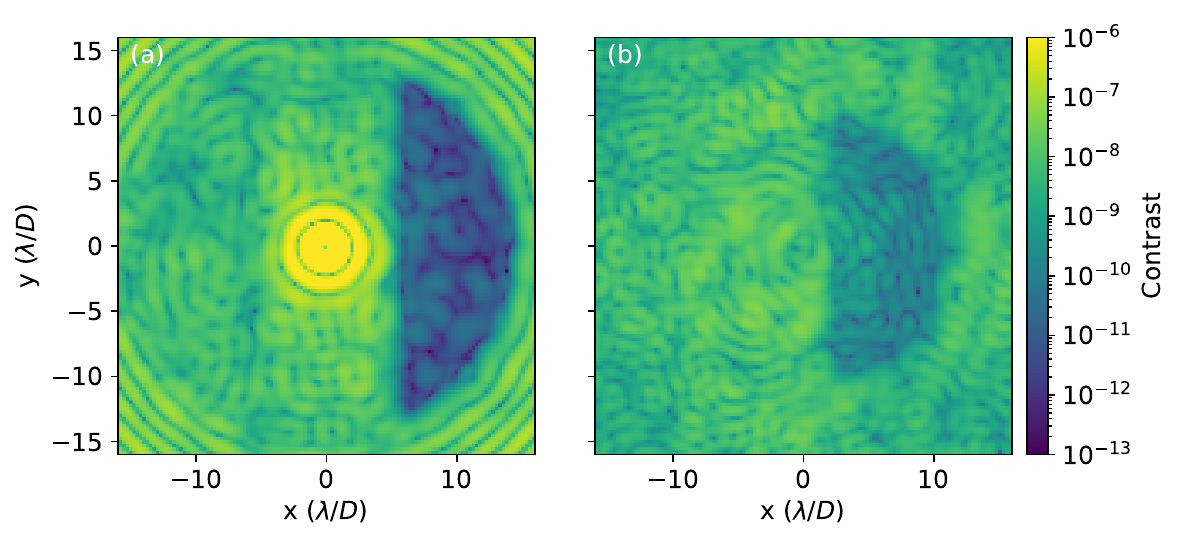}
    \caption{Simulated contrast residuals due to jitter and beam walk at \SI{300}{\nano \meter} for (a) the SPC configuration and (b) the VVC configuration. The residual is isolated by subtracting the difference of the post-EFC exit pupil fields with and without jitter and incoherently averaging the residual focal plane intensities for 50 realizations of jitter.}
    \label{fig:beamwalk_sim}
\end{figure}

The contrast floor due to jitter and beam walk is added to the model after the (jitter-free) EFC correction loop converges to the final contrast. The jitter realizations are added to the post-EFC system and incoherently summed. In integrated model runs where both jitter and DM actuator noise are included, both terms are modeled simultaneously. This allows for the possibility of coherent speckle pinning among the static speckle floor, beam walk-induced speckles, and actuator noise speckles that would otherwise not be captured by an incoherent sum of the independent effects. An example of the beam walk contrast floor in the SPC and VVC dark holes is given in Figure \ref{fig:beamwalk_sim}.

\section{Results}\label{sec:contrast-budget}

\begin{table}
\centering
\begin{tabular}{ | m{6cm} | m{3cm}| m{3cm} | m{3cm} | }
 \hline
Term & \SI{200}{\nano \meter} & \SI{300}{\nano \meter} & \SI{400}{\nano \meter} \\
 \hline
 \hline
 Chromatic EFC residuals (surface and reflectivity errors) & \makecell[l]{ 2\%: $3.4\times10^{-9}$ \\ 5\%: $1.6\times10^{-8}$ \\ 10\%: $5.6\times10^{-8}$} & \makecell[l]{ 2\%: $2.1\times10^{-9}$\\ 5\%: $8.9\times10^{-9}$ \\ 10\%:  $3.3\times10^{-8}$} & \makecell[l]{ 2\%: $2.1\times10^{-9}$ \\ 5\%: $8.4\times10^{-9}$ \\ 10\%: $3.4\times10^{-8}$ }\\
 \hline
 Mask specular reflectivity & $2.0\times10^{-10}$ & $1.4\times10^{-10}$ & $1.4\times10^{-10}$\\
 \hline
 Mask diffuse reflectivity & $4.4\times10^{-11}$ & $1.5\times10^{-10}$ & $3.3\times10^{-10}$ \\
 \hline
  Mask fabrication errors & \makecell[l]{ 2\%: $9.7\times10^{-11}$ \\ 5\%: $1.8\times10^{-10}$ \\ 10\%: $5.9\times10^{-10}$}  & \makecell[l]{ 2\%: $1.6\times10^{-10}$ \\ 5\%: $2.3\times10^{-10}$ \\ 10\%: $5.6\times10^{-10}$ }  & \makecell[l]{ 2\%: $2.2\times10^{-10}$ \\ 5\%: $3.0\times10^{-10}$ \\ 10\%: $5.2\times10^{-10}$ }  \\
 \hline
 Scattering (surface roughness) & $1.4\times10^{-11}$ & $1.4\times10^{-11}$  & $1.4\times10^{-11}$  \\
 \hline
 Jitter and beamwalk & $4.4\times10^{-11}$ & $1.6\times10^{-11}$ & $8.5\times10^{-12}$\\
 \hline
 Polarization aberrations & $1.6\times10^{-10}$ & $2.6\times10^{-10}$ & $8.2\times10^{-10}$ \\
 \hline
 DM quantization & $4.0\times10^{-11}$ & $1.8\times10^{-11}$ & $1.0\times10^{-11}$\\
 \hline
 DM noise & $7.8\times10^{-11}$ & $3.5\times10^{-11}$  & $2.0\times10^{-11}$ \\
 \hline
 \hline
 Total contrast &  \makecell[l]{ 2\%: $4.1\times10^{-9}$ \\ 5\%: $1.7\times10^{-8}$ \\ 10\%: $5.8\times10^{-8}$ } & \makecell[l]{ 2\%: $2.9\times10^{-9}$ \\ 5\%: $9.8\times10^{-9}$ \\ 10\%: $3.4\times10^{-8}$ } & \makecell[l]{ 2\%: $3.4\times10^{-9}$ \\ 5\%: $9.8\times10^{-9}$ \\ 10\%: $3.6\times10^{-8}$ }\\
 \hline
\end{tabular}
\caption 
   { \label{tab:contrast_spc} 
Contrast budget for the SPC layout for central wavelengths $\lambda_0$ from \si{200}-\SI{400}{\nano \meter}, and bandwidths from 2-10\%. The dark hole is a D-shaped, half-sided DH from $6-14\lambda_0/D$. The individual terms are incoherently added to produce the total contrast floor.}
\end{table}

\begin{table}
\centering
\begin{tabular}{ | m{6cm} | m{3cm}| m{3cm} | m{3cm} | }
 \hline
Term & \SI{200}{\nano \meter} & \SI{300}{\nano \meter} & \SI{400}{\nano \meter} \\
 \hline
 \hline
 Chromatic EFC residuals (surface and reflectivity errors) & \makecell[l]{ 2\%: $4.7\times10^{-9}$ \\ 5\%: $2.3\times10^{-8}$ \\ 10\%: $7.9\times10^{-8}$ } & \makecell[l]{ 2\%: $1.5\times10^{-9}$ \\ 5\%: $7.7\times10^{-9}$  \\ 10\%: $2.7\times10^{-8}$  } & \makecell[l]{ 2\%: $9.2\times10^{-10}$ \\ 5\%: $4.0\times10^{-9}$ \\ 10\%: $1.4\times10^{-8}$ } \\
 \hline
 Stop specular reflectivity & $2.0\times10^{-11}$ & $8.9\times10^{-12}$ & $7.9\times10^{-12}$\\
 \hline
 Stop diffuse reflectivity & $3.0\times10^{-11}$ & $8.0\times10^{-11}$ & $1.8\times10^{-10}$ \\
 \hline
 Scattering (surface roughness) & $1.2\times10^{-10}$  & $1.2\times10^{-10}$  & $1.2\times10^{-10}$  \\
 \hline
 Jitter and beamwalk & $3.6\times10^{-9}$ & $4.6\times10^{-10}$ & $2.3\times10^{-10}$\\
 \hline
 Polarization aberrations & $9.0\times10^{-13}$ & $5.3\times10^{-13}$ & $1.7\times10^{-12}$\\
 \hline
 Polarization leakage & $6.9\times10^{-11}$  & $6.3\times10^{-11}$ & $6.2\times10^{-11}$\\
 \hline
 DM quantization & $8.2\times10^{-11}$ & $3.6\times10^{-11}$ & $2.1\times10^{-11}$\\
 \hline
 DM noise & $1.6\times10^{-10}$ & $7.1\times10^{-11}$  & $4.0\times10^{-11}$ \\
 \hline
 \hline
 Total contrast & \makecell[l]{ 2\%: $9.0\times10^{-9}$ \\ 5\%: $2.7\times10^{-8}$ \\ 10\%: $8.3\times10^{-8}$ } & \makecell[l]{ 2\%: $2.3\times10^{-9}$ \\ 5\%: $8.5\times10^{-9}$ \\ 10\%: $2.8\times10^{-8}$ } & \makecell[l]{ 2\%: $1.9\times10^{-9}$ \\ 5\%: $5.0\times10^{-9}$ \\ 10\%: $1.5\times10^{-8}$ }\\
 \hline
\end{tabular}
\caption 
   { \label{tab:contrast_vvc} 
Contrast budget for the VVC layout for central wavelengths $\lambda_0$ from \si{200}-\SI{400}{\nano \meter}, and bandwidths from 2-10\%. The dark hole is a D-shaped, half-sided DH from $3-10\lambda_0/D$. The individual terms are incoherently added to produce the total contrast floor.}
\end{table}

A contrast budget that decomposes the contribution from the individual terms modeled above is given in Tables \ref{tab:contrast_spc} and \ref{tab:contrast_vvc}, under the assumption that the terms incoherently sum. As a cross-check of this budget, we integrate all these terms in an end-to-end numerical simulation. All the terms that interact coherently or can be partially corrected by EFC---surface and reflectivity errors, mask fabrication defects, DM quantization, polarization aberrations and leakage (assuming an unpolarized source)---are directly included in the Fresnel model. EFC is performed to compensate for the combined effect of these terms. Note that in the case of the SPC, we correct the mean of the $E_{xx}$ and $E_{yy}$ fields, while in the case of the VVC, we select the $E_{xx}$ field at the focal plane and correct only that term. At the conclusion of EFC, we run multiple realizations of the model with the final DM command to incorporate jitter, beamwalk, and DM actuator noise, and incoherently average the focal plane intensities. For polychromatic simulations, we incoherently average the resulting intensities for each dynamic realization and for each wavelength simulated within the bandpass. Focal plane intensities from the specular mask reflection and surface roughness scattering are simulated independently and then also incoherently added to these images. The term due to the diffuse reflectivity of the mask is assumed to be constant across the DH and is added as a bias term.

\begin{figure}
    \centering
    \includegraphics[width=1\linewidth]{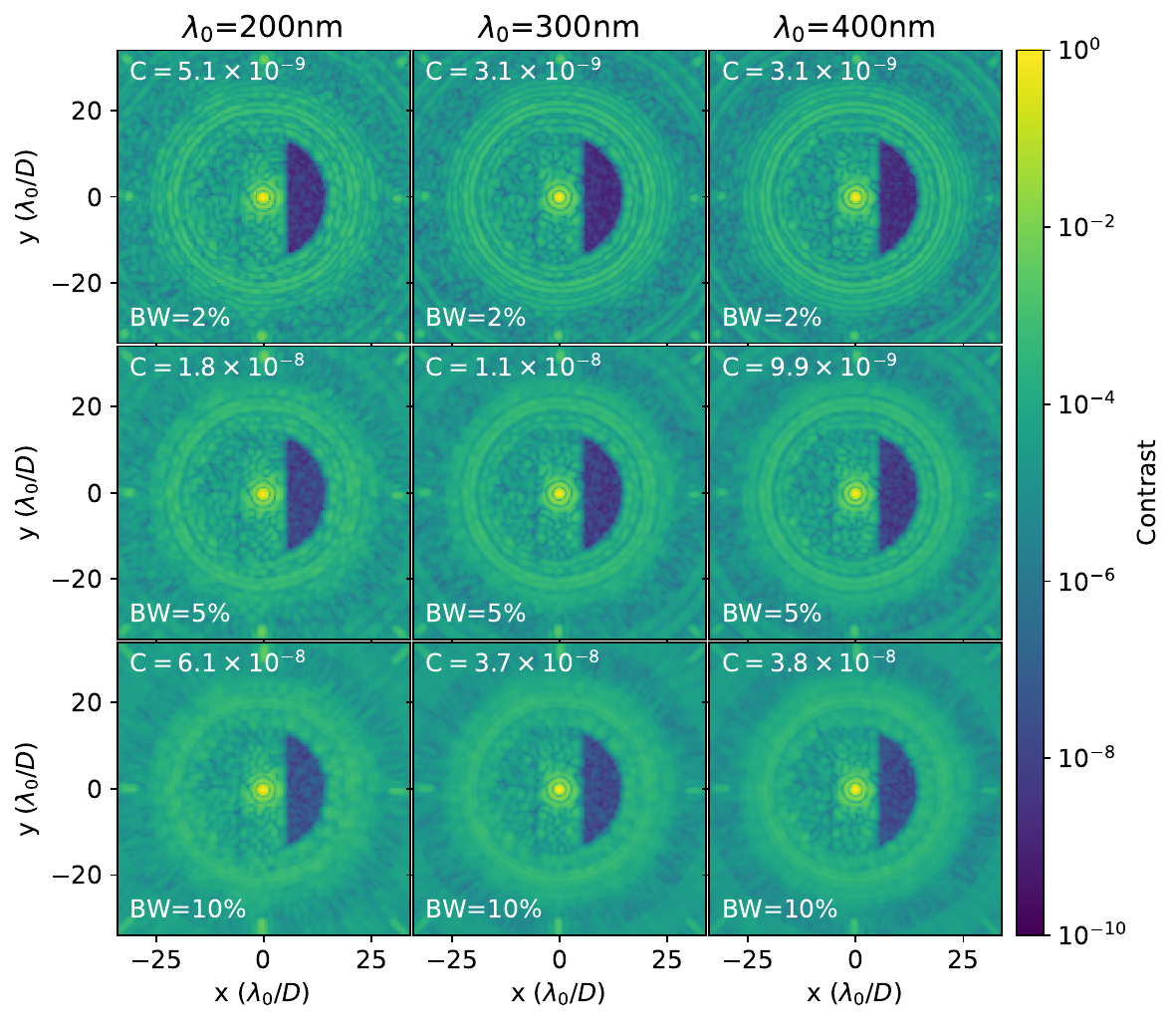}
    \caption{End-to-end SPC simulations incorporating all error terms considered here. Contrast is reported as a mean over a DH from $6-14\lambda_0/D$.}
    \label{fig:spc_integrated}
\end{figure}

\begin{figure}
    \centering
    \includegraphics[width=1\linewidth]{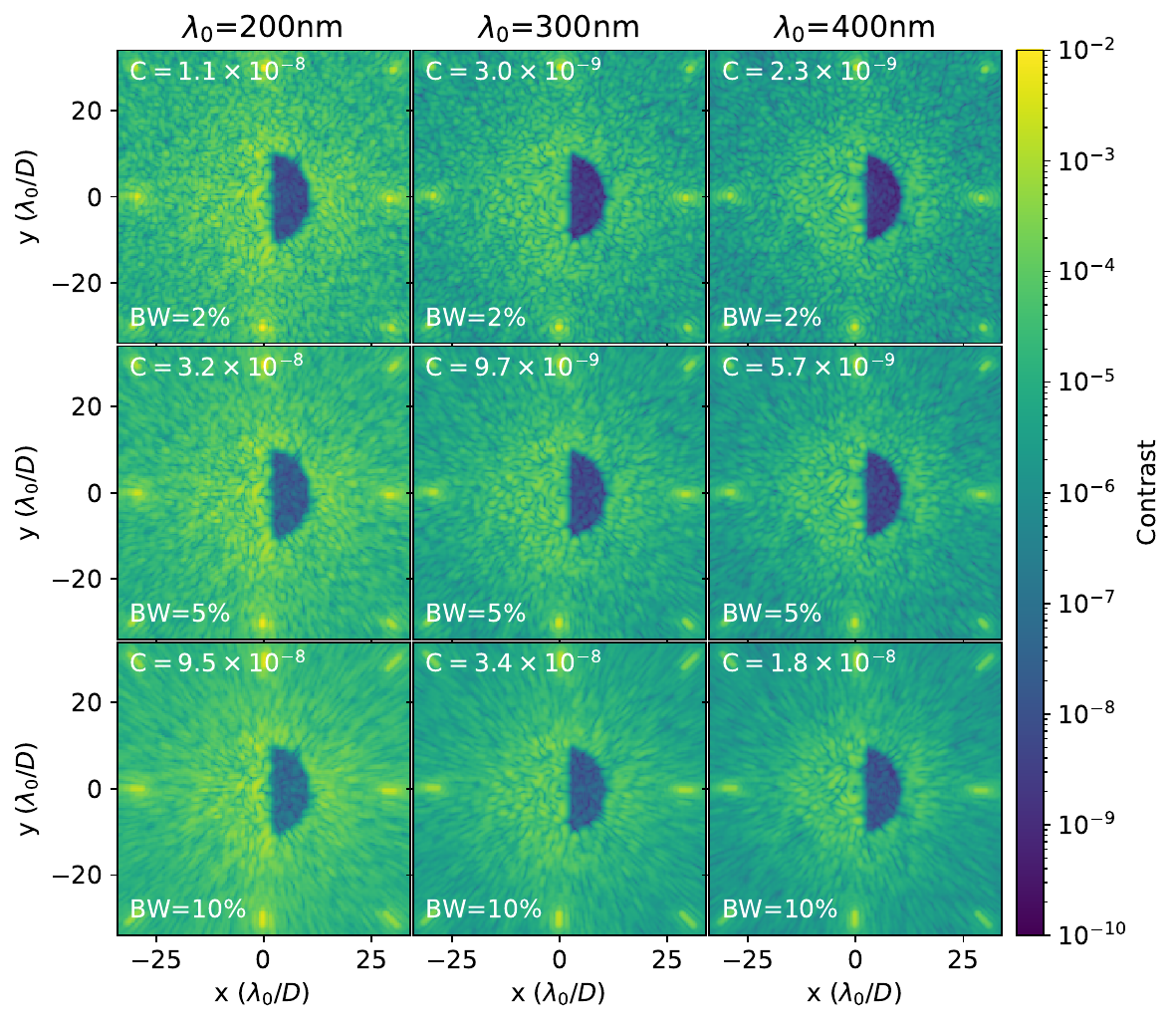}
    \caption{End-to-end VVC simulations incorporating all error terms considered here. Contrast is reported as a mean over a DH from $3-10\lambda_0/D$.}
    \label{fig:vvc_integrated}
\end{figure}

The results of these integrated simulations are shown in Figures \ref{fig:spc_integrated} and \ref{fig:vvc_integrated}, in bandwidths from 2-10\%. The mean DH contrasts from the integrated model runs closely agree with the total contrast expected from the budget, which suggests that the coherent interaction of these terms is relatively insignificant at these contrast levels.

The term that dominates the contrast budget in all cases is the chromatic EFC residual that arises from surface and reflectivity errors on out-of-pupil optics. The contrast floor from simulations of the EFC residual closely follows the $B^2$ bandwidth dependence found in Shaklan \& Green 2006\cite{Shaklan06} and later in Mazoyer \& Pueyo 2017 \cite{mazoyer_pueyo} (and found in Appendix \ref{sec:appA} as approximations of Equations \ref{eqn:contrast_bandwidth_approx} and \ref{eqn:amplitude_firstorder} in the limit of small bandwidths). Because the contrast after DM correction of amplitude and phase due to the Talbot effect is independent of the central wavelength (or very weakly dependent, in the case of amplitude) to first-order, the mitigation strategies that reduce this term for visible-wavelength coronagraphy should prove equivalently effective here: setting tighter requirements on optical surface quality, minimizing the distances of optics from conjugate pupils, and adding a second DM in a non-conjugate plane to flatten the contrast-bandwidth curve. Because Talbot lengths are longer in the UV, however, further reducing the fractional Talbot length for DH spatial frequencies with optical design choices (e.g., shifting optics closer to conjugate pupil planes or increasing the beam diameter) may run into practical limitations. EFC with two DMs in non-conjugate planes would have to be traded against the contrast penalty of adding an out-of-pupil surface with significant surface aberrations (in the case of MEMS devices, in particular) and vignetting on the second DM\cite{mazoyer_pueyo}.

\begin{figure}
    \centering
    \includegraphics[width=\linewidth]{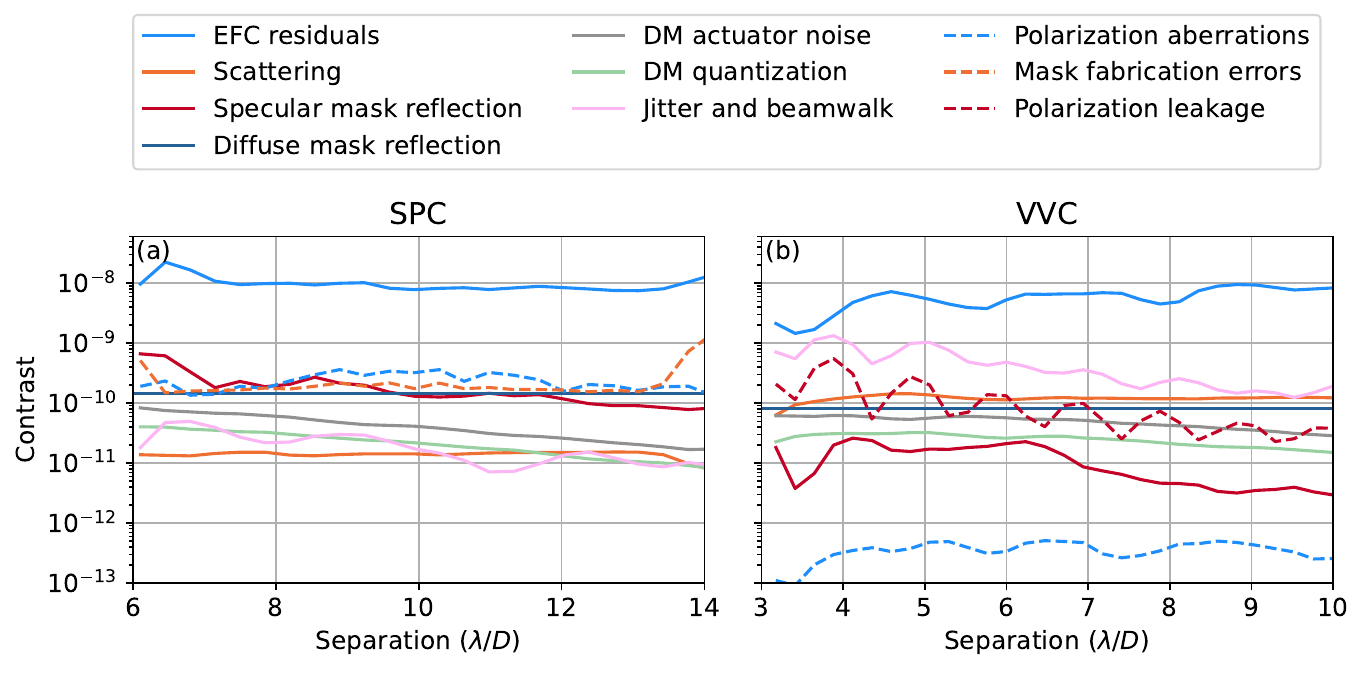}
    \caption{ Radial contrast profiles for each of the terms in Tables \ref{tab:contrast_spc} and \ref{tab:contrast_vvc} for the (a) SPC and (b) VVC configurations at $\lambda_0=\SI{300}{\nano \meter}$ and a 5\% BW. Note that the mask fabrication error term appears only in the SPC case and polarization leakage only in the VVC case.}
    \label{fig:radial_contrast}
\end{figure}

Jitter and beam walk become a significant term in the contrast budget in the VVC configuration, particularly at \SI{200}{\nano \meter}. This suggests that the current jitter levels of SCoOB in vacuum (${\leq}10^{-2} \lambda/D$ RMS at \SI{200}{\nano \meter}) may be insufficient for the shortest wavelengths in the absence of LOWFS\&C. From Equation \ref{eqn:beamwalk}, we expect the beam walk contrast to approximately scale as $C \propto (\sigma/\sigma_0)^2$, so even a modest improvement in jitter RMS $\sigma$ with LOWFS\&C can significantly lower the contrast floor. Note that the \SI{200}{\nano \meter} SPC results show a contrast almost $10\times$ lower, likely due to the combination of a coronagraph mask that is more forgiving of jitter and fewer optics in the optical train.

We plot the radial contrast profiles for the SPC and VVC in a 5\% BW centered at \SI{300}{\nano \meter} in Figure \ref{fig:radial_contrast}. The EFC chromatic residual is the largest contribution to the overall contrast at all separations and relatively flat, although the jitter and beam walk term becomes comparable in scale within $4\lambda/D$ (in the case of the VVC) in a sufficiently narrow bandwidth and at shorter central wavelengths. A modification to the optical design that lowers the EFC residual would find the contrast limited above $10^{-10}$ contrast---in the case of the VVC---by a combination of jitter and polarization leakage at smaller separations, and---in the case of the SPC---primarily by mask fabrication defects (including scattering from the BSi or CNT portions of the mask) and the shear term from polarization aberrations.

\section{Conclusions}\label{conclusions}

We have presented end-to-end simulations of two configurations of a vacuum high contrast imaging testbed operating at UV wavelengths and generated performance predictions and a contrast budget that isolates the contribution of individual physical effects to the overall contrast. In a configuration with an SPC, our model predicts a contrast of $3.0\times10^{-9}$ in a 2\% BW and $3.4\times10^{-8}$ in a 10\% BW centered at \SI{300}{\nano \meter}. In the configuration with a VVC, we predict a contrast of $2.4\times10^{-9}$ in a 2\% BW and $2.8\times10^{-9}$ in a 10\% BW at the same central wavelength. The dominant contrast limitation is the chromatic EFC residual set by surface and reflectivity errors on the optics, although a number of other terms in the budget exceed $10^{-10}$ contrast (Tables \ref{tab:contrast_spc} and \ref{tab:contrast_vvc}). This modeling effort constitutes the first step toward a demonstration of high contrast imaging in the UV in a testbed environment.

The developed model incorporates the majority of terms that have been identified in the literature as the major limitations to contrast---optical surface and reflectivity errors, scattering from surface roughness, DM quantization and actuator noise, polarization aberrations and polarization leakage, mask fabrication errors, jitter, and beam walk. There are a number of terms, however, that are not modeled here or could be refined, and may turn out to be important at UV wavelengths:
\begin{itemize}
    \item Second-order polarization aberration effects, such as those arising from spatial non-uniformity of coatings, are not modeled here. Given the relatively minor role polarization aberrations play in our contrast budget, these terms are unlikely to limit testbed performance but may be significant for a coronagraph instrument downstream of telescope optics, which typically have much stronger polarization aberrations. Additionally, while we did include bulk retardance errors, polarization optics---in this case, LPs, QWPs, and the VVC---have spatial variations of their retardance and optical axis. VVC spatial retardance variations were modeled in Anche et al. 2024\cite{anche2024} but are not included here.
    \item Our measurements of surface roughness do not extend to the $1/\lambda$ spatial frequency required for a full scattering treatment and instead rely on extrapolation from mid-spatial frequencies. Any flattening of the PSD at high spatial frequencies would lead to an underestimate of the scattering.
    \item Scattering from dust and other particulate contamination on optical surfaces is not included in this model.
    \item The reflective membranes of MEMS DMs have high spatial frequency structures at the sub-actuator scale\cite{norton_2009} that scatter light to large angles and are not fully captured by either our diffraction or scattering model.
    \item Mask features at the ${\sim} \lambda$-scale are known to induce vector diffraction effects\cite{zhang_2018} not captured in this model. Our proposed SPC mask design has a minimum feature-size of \SI{13.3}{\micro \meter}, which is likely large enough that the impact to our predicted contrast is negligible, but further study is needed to understand the significance at $10^{-10}$ contrast\cite{jenkins_2024}.
    \item The developed contrast budget does not address optical ghosts, which can impose significant limits on testbed 
    performance\cite{noyes_2023}. The SPC configuration features no transmissive optics, but the VVC configuration necessarily does. Ghosts are difficult to predict and model, but UV anti-reflection (AR) coatings with similar performance to visible-wavelength ones are readily commercially available (e.g., $\rm R{<}0.5\%$ from \SI{200}-\SI{400}{\nano \meter}), so ghosts are not expected to be any more significant in the UV.
\end{itemize}

We plan to address these limitations in our model and contrast budget, with efforts to measure OAP and DM surfaces with an atomic force microscope to probe structures to the ${\ll} \lambda$ scale, direct measurements of small angle scattering for these surfaces, measurements of polarization optics and coated surfaces with a Mueller matrix microscope with ${<}\SI{1}{\micro \meter}$ resolution to probe spatial variations of their polarization properties, and fabrication and metrology of UV-optimized coronagraph masks. These measurements of as-built optics will be incorporated into the model to update our performance predictions, and the model will continue to be refined to include higher-order terms. We note that, while SCoOB was designed to enable a demonstration of UV coronagraphy, it was not optimized for UV performance, so these contrast performance estimates should not be interpreted to be representative of the fundamental limits of high contrast imaging in the UV.


The model developed here will ultimately be validated by a laboratory demonstration of high contrast imaging at UV wavelengths in vacuum. The expectation is that lessons learned from this model development, component-level metrology, and laboratory demonstrations will help direct technology development efforts for UV coronagraphy and inform design trades for a potential HWO UV coronagraph.

\subsection*{Disclosures}
The authors declare there are no financial interests, commercial affiliations, or other potential conflicts of interest that have influenced the objectivity of this research or the writing of this paper.

\subsection*{Code, Data, and Materials Availability} 

The archived version of the simulation code developed for this manuscript can be freely accessed through Zenodo at \url{https://doi.org/10.5281/zenodo.14921084}. Other supporting materials (e.g., Jones pupils, optical surface and reflectivity maps, and SPC mask designs) can be provided upon request.

\subsection* {Acknowledgments}
Portions of this research were supported by funding from the Technology Research Initiative Fund (TRIF) of the Arizona Board of Regents
and by generous anonymous philanthropic donations to the Steward Observatory of the College of Science at the University of Arizona. This work made use of \textsc{NumPy}\cite{numpy}, \textsc{SciPy}\cite{scipy}, \textsc{Astropy}\cite{astropy}, and \textsc{HCIPy}\cite{hcipy}.

\appendix

\section{Contrast limit for a single-DM correction due to the Talbot}\label{sec:appA}

To motivate the discussion of the post-EFC chromatic contrast residuals as a function of central wavelength, we derive expressions for the contrast floor that arises from optical aberrations due to the Talbot effect. The analysis here largely follows work in Pueyo \& Kasdin 2007\cite{pueyo_polychromatic_2007} and Mazoyer \& Pueyo 2017\cite{mazoyer_pueyo}, but we provide a handful of closed-form expressions for the bandwidth-integrated contrast after DM correction in a half-sided DH, which---to the authors' knowledge---have not previously appeared in the literature.

\subsection{First-order phase correction}
In the small phase-regime, the electric field for a sinusoidal surface aberration of amplitude $\alpha$ and spatial frequency $n/D$ cycles/diameter is given by
\begin{equation}\label{eqn:efield}
    E(x) \approx 1 + i \alpha \frac{4 \pi}{\lambda} \cos\left( 2 \pi \frac{n}{D} x \right).
\end{equation}
After a Fresnel propagation to a surface a distance $z$ away, the field becomes
\begin{equation}\label{eqn:Ez_smallphase_talbot}
    E_z = 1 + i \alpha \frac{4 \pi}{\lambda} e^{-i 2 \pi z/z_T(\lambda)} \cos\left(2\pi \frac{n}{D} x \right)
\end{equation}
where we've neglected constant factors, and $z_T(\lambda)$ is the Talbot length (Equation \ref{eqn:z_talbot}). In the absence of correction, the total field (sorting the terms by the $+n \lambda/D$ and $-n \lambda/D$ speckles) becomes
\begin{equation}
\begin{aligned}
    E_z(x) = 1 + i \alpha \frac{2 \pi} {\lambda} e^{-i 2 \pi z/z_T(\lambda)}
    \Bigg\{ &
        e^{+i 2 \pi \frac{n}{D} x} + e^{-i 2 \pi \frac{n}{D} x} 
    \Bigg\}.
\end{aligned}
\end{equation}

The average contrast at the $+n \lambda/D$ location in the focal plane can be found integrating the square modulus of the coefficients modifying the $e^{+i 2 \pi \frac{n}{D} x}$ term over a flat bandwidth $\Delta \lambda$:
\begin{equation}\label{eqn:contrast_bandwidth_nocorr}
    I_{+n} =\alpha^2 \frac{4 \pi^2} {\Delta \lambda} \int_{\lambda_0 - \Delta \lambda/2}^{\lambda_0 + \Delta \lambda/2} \frac{1}{\lambda^2} d \lambda = \alpha^2 \frac{4 \pi^2} {\lambda_0^2} \left[ \frac{1}{1-\left(\frac{B}{2}\right)^2 }\right].
\end{equation}

With a small phase correction $\sigma$ on the DM, the total field of Equation \ref{eqn:Ez_smallphase_talbot} becomes approximately
\begin{align}
    E_z(x) &\approx 1 + i \alpha \frac{4 \pi} {\lambda} e^{-i 2 \pi z/z_T(\lambda)} \cos\left( 2 \pi \frac{n}{D} x \right) + i \frac{4 \pi} {\lambda} \sigma
\end{align}
For a general value of $z$, the original purely imaginary aberration term now has both a real and an imaginary component determined by the ratio $z/z_T(\lambda)$. The DM surface correction $\sigma$, however, is constrained to be purely real and specified at a single wavelength $\lambda_0$, so if we choose a DM correction of the form
\begin{equation}
    \sigma = -\alpha  \cos \left( 2 \pi \left[ \frac{n}{D} x - \frac{z}{z_T(\lambda_0)} \right]  \right),
\end{equation}
then the total field can be expressed as
\begin{equation}
\begin{aligned}
    E_z = 1 + i \alpha \frac{2 \pi} {\lambda}
    \Bigg\{ &
        e^{+i 2 \pi \frac{n}{D} x} \left[ e^{-i 2 \pi z/z_T(\lambda)} - e^{-i 2 \pi z/z_T(\lambda_0)}\right] \\
    + &   e^{-i 2 \pi \frac{n}{D} x} \left[ e^{-i 2 \pi z/z_T(\lambda)} - e^{i 2 \pi z/z_T(\lambda_0)}\right]
    \Bigg\},
\end{aligned}
\end{equation}
where it can be immediately seen that at $\lambda=\lambda_0$, the speckle at the $+n \lambda/D$ location in the focal plane is completely nulled.

In the general case for $\lambda \neq \lambda_0$, the post-coronagraph speckle intensity at the targeted $+n\lambda/D$ speckle can be found by taking the square modulus of the coefficients modifying the $e^{+i 2 \pi \frac{n}{D} x}$ term:
\begin{equation}\label{eqn:chromatic_residual}
\begin{aligned}
    I_{+n} = \alpha^2 \frac{8 \pi^2} {\lambda^2} &\left[1 - \cos\left( 2 \pi z \left[\frac{1}{z_T(\lambda)} - \frac{1}{z_T(\lambda_0)} \right] \right) \right].
\end{aligned}
\end{equation}

The contrast over a flat bandwidth $\Delta \lambda$ can be found by integrating Equation \ref{eqn:chromatic_residual}:
\begin{equation}\label{eqn:contrast_bandwidth_numerical}
\begin{aligned}
    C_\alpha
    = \alpha^2 \frac{8 \pi^2} {\Delta \lambda} \int_{\lambda_0 - \Delta \lambda/2}^{\lambda_0 + \Delta \lambda/2} \frac{1}{\lambda^2} &\left[1 - \cos\left( 2 \pi z \left[\frac{1}{z_T(\lambda)} - \frac{1}{z_T(\lambda_0)} \right] \right) \right] d\lambda.
\end{aligned}
\end{equation}

\begin{figure}
    \centering
    \includegraphics[width=0.8\linewidth]{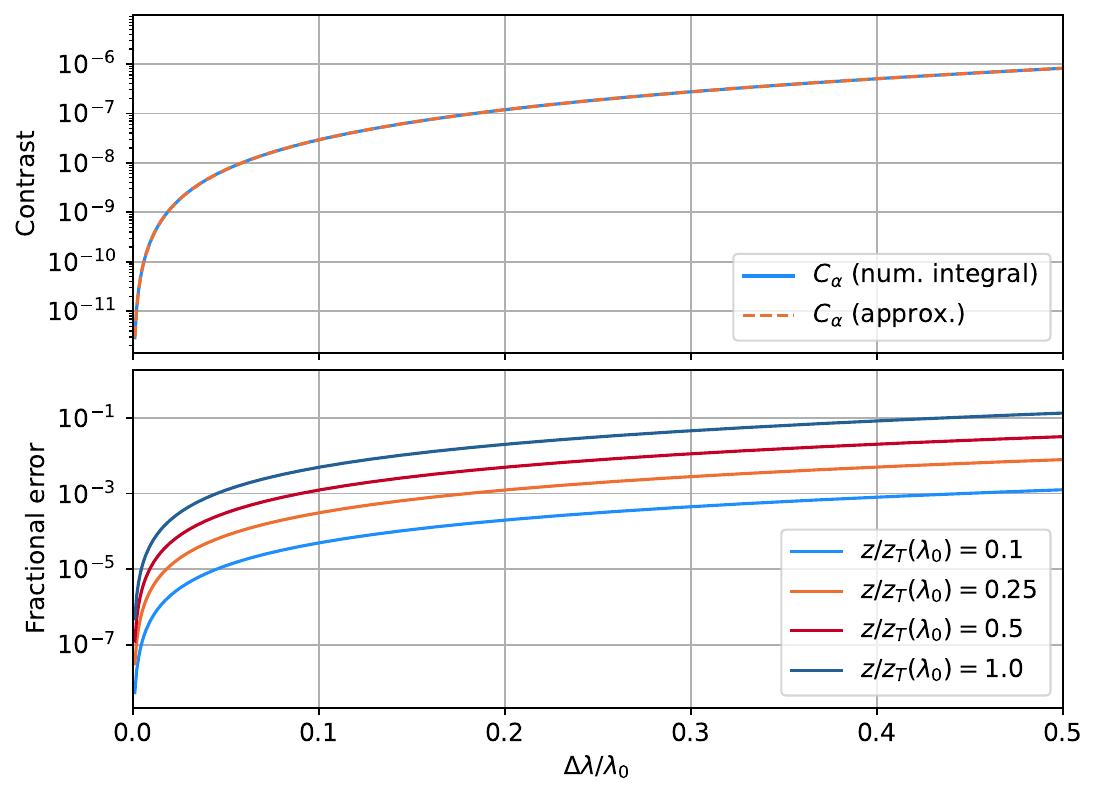}
    \caption{Top: Bandwidth-dependence of the contrast floor after DM correction of a first-order phase error in a half-sided DH. Here we assume $\alpha=\SI{1}{\nano\meter}$, $n=15$ cycles/aperture, $D=\SI{15}{\milli \meter}$, and $z=\SI{300}{\milli \meter}$. Note that this term does not depend on the central wavelength. Bottom: Comparison of the approximation in Equation \ref{eqn:contrast_bandwidth_approx} and a numerical integration of Equation \ref{eqn:contrast_bandwidth_numerical}. Residuals from the approximation remain below 1\% at a 50\% BW even up to $z=z_T(\lambda_0)/4$.}
    \label{fig:phi_firstorder}
\end{figure}

We approximate a solution to this integral by the Taylor expansion of the $\cos$ term. At the limits of the integral, we have
\begin{equation}
    \cos\left( 2 \pi z \left[\frac{1}{z_T(\lambda)} - \frac{1}{z_T(\lambda_0)} \right] \right) \rightarrow \cos \left( \pi \frac{z}{z_T(\lambda_0)} B \right),
\end{equation}
where the bandwidth $B=\Delta \lambda / \lambda_0$. This approximation will be accurate if the product of the fractional Talbot length and the bandwidth is sufficiently small. Retaining terms up to second-order, Equation \ref{eqn:contrast_bandwidth_numerical} becomes
\begin{equation}\label{eqn:contrast_bandwidth_approx}
\begin{aligned}
    C_\alpha &= \alpha^2 \frac{4 \pi^4 z^2} {\Delta \lambda} \left(\frac{n}{D}\right)^4 \int_{\lambda_0 - \Delta \lambda/2}^{\lambda_0 + \Delta \lambda/2} \left( \frac{\lambda - \lambda_0}{\lambda} \right)^2 d \lambda\\
                &= 4 \pi^4 \alpha^2 z^2 \left( \frac{n}{D} \right)^4 \left[ 1 + \frac{1}{1-(\frac{B}{2})^2} - \frac{4}{B} \tanh^{-1}\left( \frac{B}{2} \right) \right],
\end{aligned}
\end{equation}
which shows that the residual from the correction of the first-order phase term depends only on bandwidth and is independent of the central wavelength. We plot an example of this curve in Figure \ref{fig:phi_firstorder}, as well as the error of the approximation for a range of fractional Talbot lengths and bandwidths.
\subsection{First-order amplitude correction}

A field with a pure amplitude aberration of the form $E = 1 + \eta \cos \left(2 \pi (n/D) x \right)$ at a distance $z$ from the pupil plane can be monochromatically compensated in a half-sided DH with the DM correction
\begin{equation}
    \sigma = - \eta \frac{\lambda_0}{4\pi} \sin\left( 2\pi \left[ \frac{n}{D} x - \frac{z}{z_T(\lambda_0)} \right]  \right).
\end{equation}
Repeating the analysis in the previous section yields a similar expression for the bandwidth-integrated contrast
\begin{equation}\label{eqn:amplitude_firstorder}
    C_\eta = \frac{\eta^2}{4} \left[1 - 2 \lambda_0^2 \pi^2 z^2 \left( \frac{n}{D}\right)^4 + \frac{1}{\left[1-\left(\frac{B}{2}\right)^2 \right]} - \frac{4}{B} \left( 1 - \lambda_0^2 \pi^2 z^2 \left( \frac{n}{D}\right)^4 \right) \tanh^{-1}\left(\frac{B}{2}\right) \right].
\end{equation}

\begin{figure}
    \centering
    \includegraphics[width=0.8\linewidth]{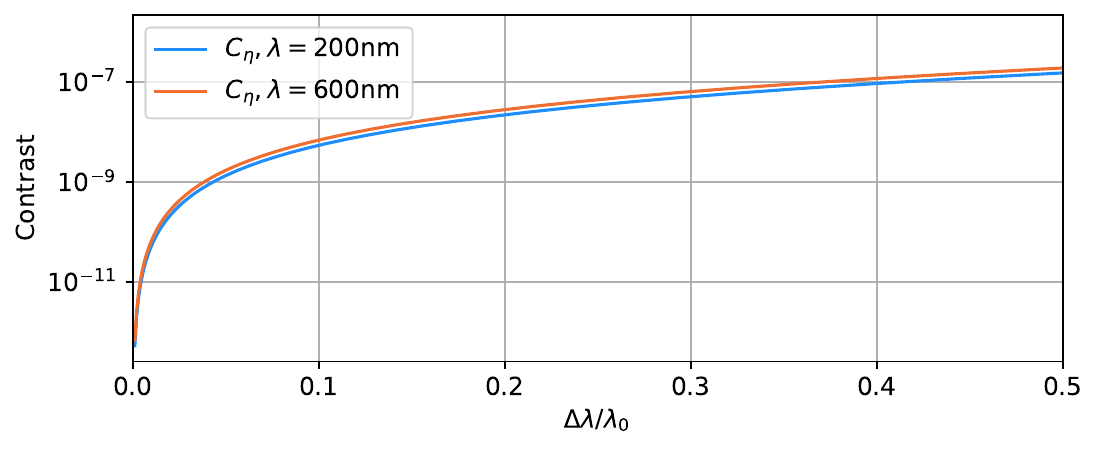}
    \caption{Bandwidth-dependence of the contrast floor after DM correction of a pure amplitude aberration in a half-sided DH. Here we assume $\eta=0.5\%$, $n=15$ cycles/aperture, $D=\SI{15}{\milli \meter}$, and $z=\SI{300}{\milli \meter}$. We plot at $\lambda_0=\SI{200}{\nano \meter}$ and \SI{600}{\nano \meter} to show the weak dependence on central wavelength.}
    \label{fig:amp_firstorder}
\end{figure}

Note that the additional terms that depend on Talbot length and central wavelength are scaled by a $\lambda_0^2 \pi^2 z^2 \left( \frac{n}{D}\right)^4$ factor compared to the wavelength-independent terms. For a typical coronagraph instrument design, this factor is $\lessapprox 10^{-2}$ and becomes smaller with wavelength. We plot an example of the contrast residual after DM correction for a fixed amplitude aberration at \SI{200}{\nano \meter} and \SI{600}{\nano \meter} in Figure \ref{fig:amp_firstorder}, which shows only a weak dependence on the central wavelength (and, in fact, shows a marginal improvement in the contrast at \SI{200}{\nano \meter}).

\subsection{Second-order phase correction}

The 2nd-order term from a cosine phase aberration of amplitude $\beta$ at spatial frequency $\frac{n}{2D}$ will contribute an electric field of the form
\begin{equation}
    E_z = - \beta^2 \frac{4 \pi^2}{\lambda^2}\cos\left( 2 \pi \frac{n}{D} x \right) e^{-i 2 \pi z/z_T(\lambda)}
\end{equation}
at the $n/D$ spatial frequency. With a single-DM correction for a half-sided DH (ignoring the 2nd-order term on the DM) of the form
\begin{equation}
    \sigma = \beta^2 \frac{\pi} {\lambda_0}  \sin \left( 2 \pi \left[ \frac{n}{D} x - \frac{z}{z_T(\lambda_0)} \right]  \right),
\end{equation}
the residual contrast at the $+n \lambda/D$ speckle integrated over the bandwidth will be
\begin{equation}
    C_\beta = 4 \pi^4 \frac{\beta^4} {\Delta \lambda} \int_{\lambda_0 - \Delta \lambda/2}^{\lambda_0 + \Delta \lambda/2}
              \left[ \frac{1}{\lambda_0^2 \lambda^2} + \frac{1}{\lambda^4} -\frac{2}{\lambda^3 \lambda_0} \cos \left( 2 \pi z \left( \frac{1}{z_T(\lambda)} - \frac{1}{z_T(\lambda_0)} \right) \right) \right] d \lambda.
\end{equation}
Retaining terms up to the second-order in the Taylor expansion for $\cos$, this integral evaluates to:
\begin{equation}\label{eqn:contrast_secondorder}
\begin{aligned}
    C_\beta \approx 4 \pi^4 \beta^4 \Bigg\{ \frac{1}{\lambda_0^4} &\left[
        \frac{1} {1-\left(\frac{B}{2}\right)^2}
        - \frac{1}{3B} \left[ \frac{1}{(1 + \frac{B}{2})^3} - \frac{1}{(1 - \frac{B}{2})^3} \right]
        - \frac{2}{\left(1-\left(\frac{B}{2}\right)^2 \right)^2} \right]  \\
        + \frac{\pi^2 z^2 \left( \frac{n}{D} \right)^4}{\lambda_0^2} &\left[\frac{2}{B} \tanh^{-1}\left(\frac{B}{2}\right) - \frac{1 - \frac{B^2}{2}} {\left( 1 - \left(\frac{B}{2} \right)^2 \right)^2}
        \right] \Bigg\}
\end{aligned}
\end{equation}

\begin{figure}
    \centering
    \includegraphics[width=0.8\linewidth]{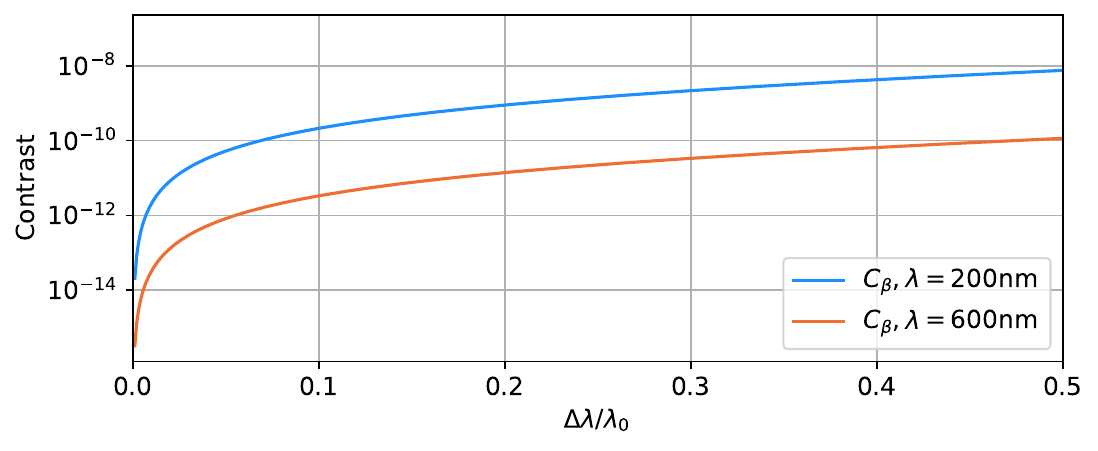}
    \caption{Bandwidth-dependence of the contrast floor at $n \lambda/D$ after DM correction in a half-sided DH of the second-order term of a sinusoidal phase aberration at $n/2$ cycles/aperture. Here we assume $\beta=\SI{1}{\nano\meter}$, $n=15$ cycles/aperture, $D=\SI{15}{\milli \meter}$, and $z=\SI{300}{\milli \meter}$. We plot at $\lambda_0=\SI{200}{\nano \meter}$ and \SI{600}{\nano \meter} to show the strong dependence on central wavelength.}
    \label{fig:phi_secondorder}
\end{figure}

Note that this second-order correction has a $\lambda_0^{-4}$ dependence on the central wavelength. As with the first-order amplitude aberration case, the terms that show a Talbot length dependence are scaled by a factor of $\pi^2 z^2 \left( \frac{n}{D} \right)^4 \lambda_0^2$, which tends to be very small. We compare the contrast residual after DM correction for a fixed second-order phase aberration at \SI{200}{\nano \meter} and \SI{600}{\nano \meter} in Figure \ref{fig:phi_secondorder}, which illustrates the strong dependence of this term on the central wavelength.

\section{Contrast limit for beam walk on a single surface}\label{sec:beamwalk_deriv}

If we introduce a small phase shift $\Delta \phi$ on the surface aberration defined in Equation \ref{eqn:efield} but retain the DM command that corrects the aberration prior to the phase shift, the electric field at $\lambda=\lambda_0$ for the corrected speckle in the half-DH becomes
\begin{equation}
    E_{+n} = i \alpha \frac{2 \pi} {\lambda_0} \left[ e^{i \Delta \phi} e^{-i 2 \pi z/z_T(\lambda_0)} - e^{-i 2 \pi z/z_T(\lambda_0)}\right]
\end{equation}
with normalized intensity
\begin{equation}
    I_{+n} = \alpha^2 \frac{4 \pi^2} {\lambda_0^2} \sin^2\left( \frac{\Delta \phi} {2}\right).
\end{equation}

For a small tip/tilt angle $\gamma$ in the pupil plane, the phase shift on the sinusoidal surface aberration a distance $z$ away is $\Delta \phi = 2\pi \gamma z \frac{n}{D}$. If we assume a Gaussian distribution of jitter angles with standard deviation $\sigma$, the contrast due to the incoherent addition of these terms is
\begin{equation}
\begin{aligned}
    C &= \alpha^2 \frac{4 \pi^2} {\lambda_0^2} \frac{1}{\sqrt{2 \pi \sigma^2}} \int_{-\infty}^{+\infty} \sin^2\left( \pi z \frac{n}{D} \gamma \right) e^{-\frac{\gamma^2}{2\sigma^2}} d \gamma \\
     &= \alpha^2 \frac{2 \pi^2}{\lambda_0^2} \left[1 - e^{-2\pi^2 \left( z \frac{n}{D} \sigma \right)^2} \right].
\end{aligned}
\end{equation}

\section{Jones pupils for VVC configuration}
\label{appendix:jones}
\begin{figure}[!ht]
    \centering
    \includegraphics[width=1\linewidth]{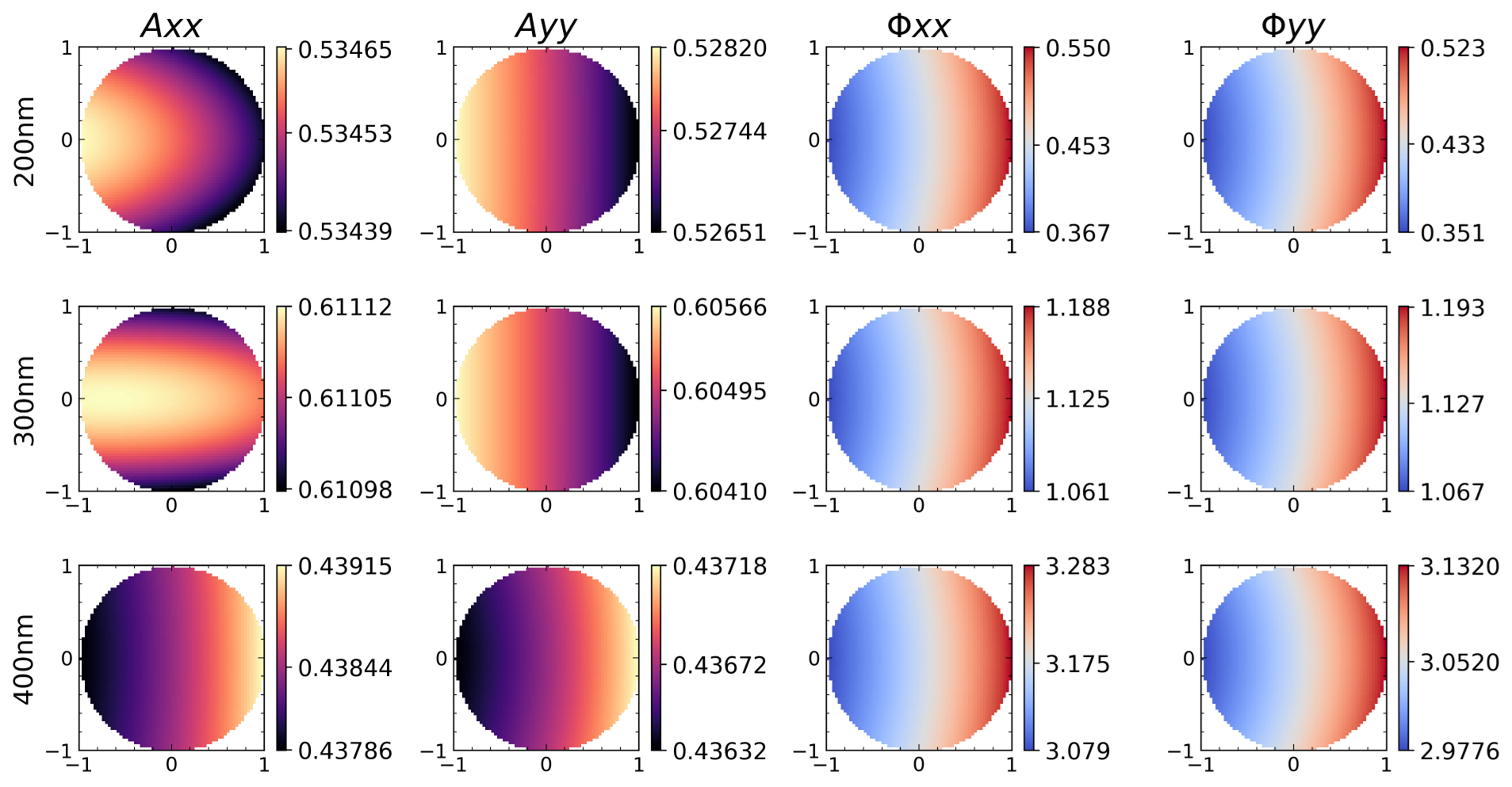}
    \caption{Jones amplitude and phase maps showing the diagonal terms estimated in the exit pupil of the VVC configuration}
    \label{fig:jones-diag-vvc}
\end{figure}
\begin{figure}[!ht]
    \centering
    \includegraphics[width=1\linewidth]{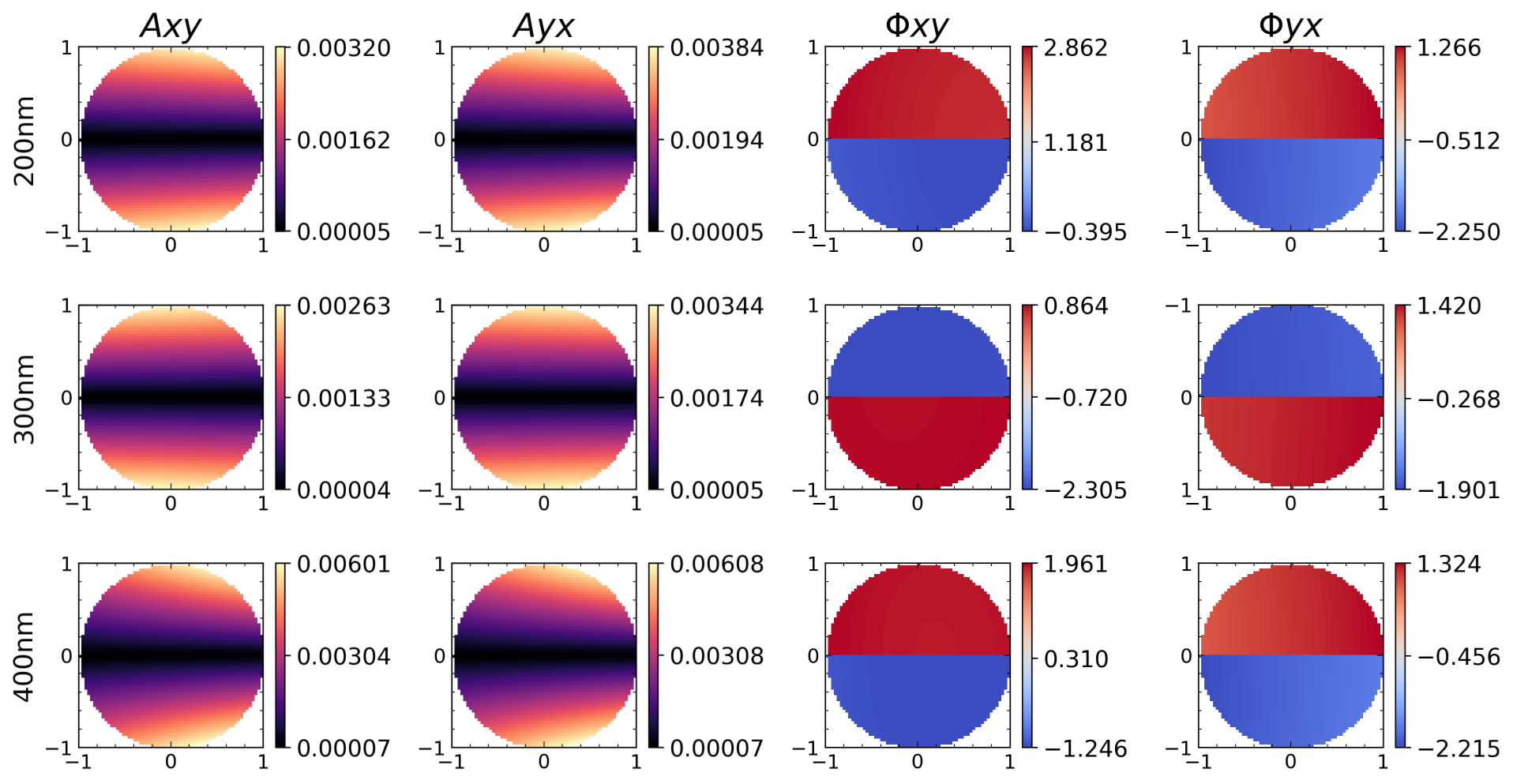}
    \caption{Jones amplitude and phase maps showing the crossed terms estimated in the exit pupil of the VVC configuration}
    \label{fig:jones-crossed-vvc}
\end{figure}

\bibliography{report}   
\bibliographystyle{spiejour}   



\vspace{1ex}

\noindent Biographies and photographs of the authors are not available.

\listoffigures
\listoftables

\end{document}